\documentclass[%
reprint,amsmath,amssymb,aps,prx]{revtex4-1}

\usepackage{graphicx}
\usepackage{dcolumn}
\usepackage{subcaption}
\usepackage{bm}
\usepackage[title]{appendix}

\def\d{\,\mathrm{d}}

\begin{document}
	
	
	\title{Spatial patterns emerging from a stochastic process near criticality}
	
	\author{Fabio Peruzzo}
	\author{Mauro Mobilia}%
	\author{Sandro Azaele}%
	\email{s.azaele@leeds.ac.uk}
	\thanks{Current address: Dipartimento di Fisica e Astronomia "Galileo Galilei", Universit\`a di Padova, Via Marzolo 8, 35131 Padova, Italy.\\ Current email: sandro.azaele@unipd.it}
	\affiliation{%
		Department of Applied Mathematics, University of Leeds
	}%
	

	\begin{abstract}
		There is mounting empirical evidence that many communities of living organisms display key features which closely resemble those of physical systems at criticality. We here introduce a minimal model framework for the dynamics of a community of individuals which undergoes local birth-death, immigration and local jumps on a regular lattice. We study its properties when the system is close to its critical point. Even if this model violates detailed balance, within a physically relevant regime dominated by fluctuations, it is possible to calculate analytically the probability density function of the number of individuals living in a given volume, which captures the close-to-critical behavior of the community across spatial scales. We find that the resulting distribution satisfies an equation where spatial effects are encoded in appropriate functions of space, which we calculate explicitly. The validity of the analytical formul\ae{} is confirmed by simulations in the expected regimes. We finally discuss how this model in the critical-like regime is in agreement with several biodiversity patterns observed in tropical rain forests. 
		
	\end{abstract}

	\maketitle
	
\section{\label{}Introduction}
Several authors have showed that the parameters of models which describe biological systems are not located at random in their parameter space, but are preferably poised in the vicinity of a point or surface which separates regimes of qualitatively different behaviors \cite{mora2011biological}. In this sense, stationary states of living systems are not only far from equilibrium, but bring the hallmark of criticality. Although the connection between underpinning dynamics and measurable quantities is sometimes tenuous and hence conclusions about criticality loose, empirical examples span a wide range of biological organization, from gene expression in macrophage dynamics \cite{nykter2008gene}, to cell growth \cite{furusawa2012adaptation}, relatively small networks of neurons \cite{schneidman2006weak}, flocks of birds \cite{cavagna2010scale} and, possibly, tree populations in tropical forests \cite{tovo2017upscaling}.
	
In this article, we mainly focus on the spatial patterns emerging from a minimal model of population dynamics close to its critical point. This latter is identified as a singularity in the population size of the system, similarly to what happens in the theory of branching processes in the sub-critical regime \cite{athreya2004branching}, when the fluctuations play a crucial role. Therefore, in our model the critical point does not mark a transition between ordered and disordered phases \textit{sensu} equilibrium statistical mechanics \cite{cardy1996scaling}, although connections in a broader context may certainly exist. The emergent patterns are not calculated by using classical size-expansion methods, but introducing a parameter expansion which appropriately identifies criticality in the parameter space of the model.
	 
The calculation of the probability distribution of large scale configurations emerging from the microscopic dynamics is challenging, even at stationarity \cite{jahnke2007solving,mellor2016characterization}. When stochastic processes violate detailed balance, they have a generator which is not self-adjoint \cite{pavliotis2014stochastic} and different states are coupled by probability currents at microscopic level \cite{zia2007probability}. These flows of probability among microstates break detailed balance, time symmetry and produce macroscopic non-equilibrium behavior. A common way to overcome these hurdles is to formulate some kind of effective Langevin equation which describes the dynamics of the mesoscopic variables of interest, loosing track, however, of the underlying microscopic dynamics \cite{henkel2008non,garcia2012noise,shem2017solution}. Nonetheless, in this paper we study a model which, despite violating microscopic detailed balance \cite{krapivsky2010kinetic,grilli2012absence}, allows one to study analytically (stationary) out-of-equilibrium properties of spatial patterns. These latter emerge mainly because of the large intrinsic fluctuations of the local population sizes. Also, the model's mathematical amenability allows us to analyze in detail those spatial ecological patterns and to compare them with observation data for ecosystems with large species richness. The agreement between model predictions and empirical data highlights the usefulness of the approach and strengthens the connections between physics and theoretical ecology.

In this spatial metacommunity model, local communities (or, equivalently, sites or voxels) are located on a $d$-dimensional regular graph (or lattice) where individuals are treated as well-mixed particles which undergo a birth and death process with local diffusion and constant colonization. We thoroughly investigate the spatial stochastic dynamics close to criticality and deduce the equation governing the evolution of the conditional distribution $ p(N|V;t) $, the probability to find $ N $ individuals in a volume $ V $ at time $ t $ (in dimension $ d $). Within this regime we map the equation of $ p(N|V) $ of the out-of-equilibrium spatial model into an equation of a suitable stochastic process, which instead satisfies detailed balance. This model is described by functions of space, which we are able to calculate exactly. The exact stochastic simulations are always matched by our analytical formul\ae{} in the expected regimes.

The rest of the paper is organized as follows: in section II we introduce the master equation of the model; in section III we calculate the mean and pairwise correlation; in section IV we study the generating function of the conditional probability density function of $ N(V) $; in section V and VI we calculate the population variance and the conditional pdf of $ N(V) $, respectively, along with a comparison between simulations and analytical formul\ae{}; in section VII we present an ecological application of the model; finally, section VIII includes some discussions and perspectives about the results.

\section{\label{ME_model_par}Master equation of the model}
\begin{figure*}[t]
		\includegraphics[width=\textwidth]{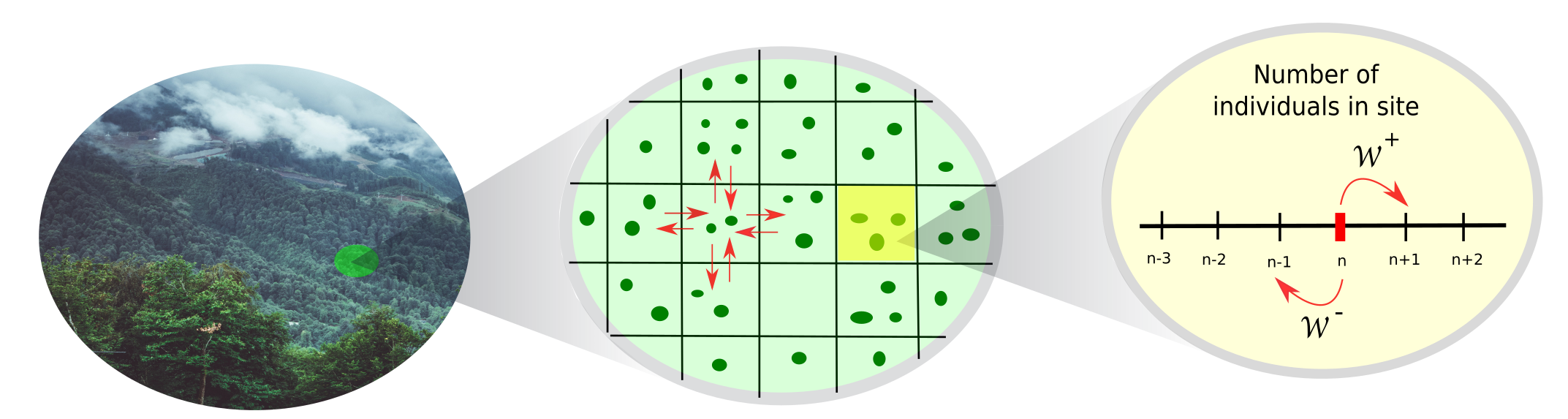}
		\caption{\textbf{Illustration of the model.} Individual trees are represented by dark green circles within local communities which are located on a regular graph (lattice). Each individual may die or give birth to an offspring with constant per capita rates. New individuals may either remain in the  community where they were born with probability $\gamma$, or hop onto one of the 2$d$ nearest neighbours with probability $1-\gamma$. Finally, all communities are colonized by external individuals at a constant immigration rate $ b_0 $. The dynamics of the model is therefore defined by the jump rates defined in eq.(\ref{rates}).}
		\label{Model_cartoon}
\end{figure*}
	
	This is a metapopulation model in which individuals live in local communities (or sites) located on a $d$-dimensional lattice, $\mathbb{L}$, whose linear side is $a$. If $X_i$, $i \in \mathbb{L}$, indicates an individual living in site $i$, the reactions defining the model's dynamics can be cast into the form
	\begin{align*}
	X_i \xrightarrow{b\gamma}& \ 2 X_i\\
	X_i \xrightarrow{\frac{b(1-\gamma)}{2d}}& \ X_i+X_j\\
	X_i \xrightarrow{r}& \ \emptyset\\
	\emptyset \xrightarrow{b_0}& \ X_i
	\end{align*}
	where $j$ indicates a site which is a nearest neighbor of the site $i$. In this model, individuals within local communities (or, equivalently, sites or voxels) are treated as diluted, well-mixed point-like particles which undergo a minimal stochastic demographic dynamics: each individual may die at a constant death rate $r$ and give birth to an offspring at a constant rate $b$. The newborn individual remains in the same community with probability $\gamma$, whereas it may hop onto one of the 2$d$ nearest neighbours with probability $1-\gamma$. Also, all communities are colonized by external individuals at a constant immigration rate $b_0$, which prevents the system to end up into an absorbing state without individuals \cite{o2010field,o2018cross}. 
	
	Notice that (for $ 0\leq\gamma<1 $) spatial movement is always coupled to birth, so that only newborn individuals can move. This is because we have in mind an application to spatial ecology, where this model mimics the population dynamics in species-rich communities of trees, in which only seeds can move. However, it can be easily modified to include random walk behaviour or different dispersals -- like those for bacteria or humans -- in which the local birth and the hopping probability are in general decoupled. We have indeed verified that the generality of our final results does not depend on that coupling (see Appendix \ref{AppF}).
	
	In the language of chemical reaction kinetics, the first reaction represents an autocatalytic production; this and the hopping move are responsible for the break of detailed balance as shown in Appendix \ref{BDB}. Therefore, stationary states of this process are non-equilibrium steady states, albeit the model is defined by linear birth and death rates.
	
Indeed, it is worth emphasizing that each ``patch'' has not a maximum number of individuals which it can accommodate, but any population size is allowed, albeit large sizes have an exponentially small probability to occur. This is because the model has no intrinsic carrying capacity which leads the population to saturation. A carrying capacity has the advantage to bring in more realistic features, but it also makes the model mathematically more complicated because of non-linear terms. Here we show that linear rates of a stochastic model \textit{per se} can produce a relevant phenomenology within a stationary out-of-equilibrium model. Thus, since we wanted to focus on the regime near criticality, we have preferred to delve into a relatively simpler system, in which non-linearities are neglected in a first approximation. More complicated dynamics are definitely important and will be studied in future works.

Finally, for the system to avert demographic explosion we have to assume that $b_0>0$ and $0<b<r$, but it turns out that the most interesting features emerge when $ b\simeq r $, i.e., close to its critical point. Indeed, as we will show in Section \ref{Eco}, for comparable birth and death rates the model is able to describe several spatial patterns of tree species in tropical forests \cite{azaele2016statistical}.

	
	
	Let us now indicate with $n_i$ the number of individuals in site $i$.
	Assuming that within every site the spatial structure can be neglected and that we have perfect mixing, when the configuration of the system is $\{n\}=\{n_i: i \in \mathbb{L}\}$, the linear birth and death rates in site $ i $, i.e., $\mathcal{W}^{+}_i(\{n\})$ and $\mathcal{W}^{-}_i(\{n\})$, read respectively
	\label{jump_rates}
	\begin{align} 
	\nonumber
	\mathcal{W}^{+}_i(\{n\})= & \frac{b(1-\gamma)}{2 d} \sum_{j: |j-i|=1} n_j +\ b \gamma n_i +b_0 \\
	\label{rates}
	\mathcal{W}^-_i(\{n\})=& \ r n_i\quad.
	\end{align}
	Let $P(\{n\},t)$ be the probability to find the system in the configuration $ \{n\} $ at time $ t $. Then the master equation for $P(\{n\},t)$ reads
	
	\begin{align}
	\label{ME}
	\frac{\partial}{\partial t} P(\{n\},t)=&\sum_{i \in \mathbb{L}}\Big\{ \\
	\nonumber
	&\mathcal{W}_i^+(\{..., n_i-1, ...\}) P(\{..., n_i-1, ...\},t)+\\
	\nonumber
	-& \mathcal{W}_i^+(\{n\}) P(\{n\},t)+\\
	\nonumber
	+& \mathcal{W}_i^-(\{..., n_i+1, ...\}) P(\{..., n_i+1, ...\},t)+\\
	\nonumber
	-&\mathcal{W}_i^-(\{n\}) P(\{n\},t) \Big\}
	\end{align}
	where the dots represent that all other occupation numbers remain as in $\{ n \}$ and it is intended that $P(\cdot)=0$ whenever any of the entrances is negative. The spatial generating function of the process is defined as
	\begin{align}
	\nonumber
	\zeta(\{H\},t)=&\langle e^{\sum_{k\in \mathbb{L}}n_{k} H_{k}} \rangle=\\
	\label{gen_func_def}
	=&\sum_{\{n\}} e^{\sum_{k \in \mathbb{L}}n_{k} H_{k}} p(\{n\},t)
	\end{align}
	where $H_i \leq 0$ for every $i \in \mathbb{L}$. Multiplying both sides of eq.(\ref{ME}) by $e^{\sum_{k' \in \mathbb{L}} n_{k'} H_{k'}}$ and summing over all configurations of the system, we find that $\zeta(\{H\},t)$ satisfies the equation
	\begin{align}
	\label{eqzeta_full}
	\frac{\partial}{ \partial t} \zeta(\{H\},t)=&\sum_{i\in \mathbb{L}} \Big\{ (e^{H_i}-1) \Big[\frac{b(1-\gamma)}{2d} \sum_{j:\mid i-j \mid=1} \frac{\partial \zeta}{\partial H_j} \\
	&+b \gamma \frac{\partial \zeta}{\partial H_i}+b_0 \ \zeta \Big]+r(e^{-H_i}-1)\frac{\partial \zeta}{\partial H_i} \Big\}\;. \nonumber
	\end{align}
	This is the main equation of the model from which we will calculate the most important results. We were not able to find the full solution of this equation. However, one can gain a lot of information about the general properties of the process by looking into the probability distribution of the random variable $N(V,t)= \sum_{i \in \mathcal{V}} n_i(t)$, where $\mathcal{V}$ is the set of sites in a $ d $-dim volume. Before studying such a distribution, it is useful to calculate the mean number of individuals and the spatial correlation between any pair of sites.
	
	\section{\label{Mean_and_pair_section} Mean and Pair Correlation}
	The equation for the mean number of individuals in the site $ k $ can be obtained by taking the partial derivative of both sides of eq.(\ref{eqzeta_full}) with respect to $H_k$ and then setting $\{H\}=0$:
	\[
	\frac{\partial \langle n_k \rangle}{\partial t}= \frac{b(1-\gamma)}{2 d} \Delta_k \langle n_k \rangle -\mu \langle n_k \rangle +b_0
	\]
	where $\mu:= r-b$ and $\Delta_k$ is the discrete Laplace operator, which is defined as
	\[
	\Delta_k f(k)=\sum_{j: \mid k-j \mid=1} \Big(f(j)- f(k) \Big)\quad.
	\]
	This finite difference equation can be solved in full generality and at stationarity, i.e. for $t \to \infty$, we get simply $\langle n \rangle=\frac{b_0}{\mu}$, regardless of any spatial location.
	
	The pairwise spatial correlation between sites $ k $ and $ l $, i.e., $\langle n_k n_l \rangle$, can also be obtained by taking the partial derivatives of both sides of eq.(\ref{eqzeta_full}) with respect to $H_k$ and $H_l$, and then setting $\{H\}=0$ (see Appendix \ref{App_A} for details): 
	\label{eq:pcf-discrete}
	\begin{align} 
	\nonumber 
	\frac{\partial}{\partial t} \langle n_k n_l \rangle=& D \Big( \Delta_k \langle n_k n_l \rangle + \Delta_l \langle n_k n_l \rangle \Big)+\\ 
	\label{pp_equation}
	-&2 \mu \langle n_k n_l \rangle +2 b_0 \langle n \rangle+\\
	\nonumber
	+&\delta_{k,l} \Big( 2 \sigma^2 \langle n \rangle +b_0 + D \Delta_k \langle n_k\rangle\Big)
	\end{align}
	where we have used the notation 
	\begin{align*}
	D:=\frac{b(1-\gamma)}{2 d} \qquad \text{and}\qquad
	\sigma^2:=\frac{r+b}{2}\quad .
	\end{align*}
	We also introduce 
	\begin{align*}
	\lambda:=\sqrt{\frac{D}{\mu}} \qquad \text{and}\qquad \rho:=\sqrt{\frac{\sigma^2}{b_0}}\quad,
	\end{align*}
	which are dimensionless parameters and provide important information about how spatial diffusion intermingles with demographic dynamics.
	
	In order to solve eq.(\ref{pp_equation}), we introduce a $ d $-dim system of Cartesian coordinates where the coordinates of each site are given as a multiple of the lattice side $ a $. Thus, a vector $ \textbf{k} $ indicates the corresponding position of a site. In this way, we can calculate the stationary solution of eq.(\ref{pp_equation}) by writing  $\langle n_{\textbf{k}} n_{\textbf{l}} \rangle$ as a Fourier series expansion. Exploiting translation invariance the final expression of the solution reads (see Appendix \ref{App_A})
	\begin{align}
	\nonumber
	\langle n_{\textbf{k}} n_{\textbf{l}} \rangle&= \langle n \rangle^2 + \langle n \rangle^2 \rho^2 \Big( 1+\frac{\mu}{2 \sigma^2}\Big) \times\\ 
	\label{eq:corr_discrete}
	&\times \Big( \frac{a}{2 \pi} \Big)^d \int_\mathcal{C} \d \textbf{p} \frac{e^{i \textbf{p} \cdot (\textbf{k}-\textbf{l})}}{1+2\lambda^2 \sum_{i=1}^d (1-\cos(p_i a))}\\ 
	\nonumber
	\end{align}
	where $p_i$ is the $ i $-th Cartesian component of the $ d $-dim vector $\textbf{p}$ and $\mathcal{C}$ is the hypercubic primitive unit cell of size $2 \pi/a$. Interestingly, pairs of sites de-correlate for $ \gamma=1 $ or $ b=0 $, when at stationarity we obtain $ \langle n_{\textbf{k}} n_{\textbf{l}} \rangle - \langle n \rangle^2 =c\delta_{\textbf{k},\textbf{l}} $, where $ c $ is a constant depending only on the demographic parameters and $ \delta_{\textbf{k},\textbf{l}} $ is a Kronecker delta. However, $ c\neq \langle n \rangle $, pointing out that local fluctuations are non-Poissonian.
	
	Eq.(\ref{eq:corr_discrete}) is amenable to a continuous spatial limit, obtained as $a \to 0$, and provides a closed analytic form for the pair correlation. Indicating now with $n(x)$ and $n(y)$ the density of individuals on the sites located at $x$ and $y$, respectively, in continuous space (and rescaling parameters accordingly), we find (see Appendix \ref{App_A})
	\begin{align}
	\label{pp_correlation}
	\frac{\langle n(x) n(y) \rangle}{\langle n \rangle^2}= 1+ &\frac{\bar{\rho}^2}{(2 \pi \bar{\lambda}^2)^{d/2}} \Big( 1+\frac{\mu}{2 \sigma^2} \Big) \times\\
	\nonumber
	\times & \Big( \frac{|x-y|}{\bar{\lambda}} \Big)^{\frac{2-d}{2}} K_{\frac{2-d}{2}} \Big(\frac{|x-y|}{ \bar{\lambda}} \Big)
	\end{align}
	where $|x-y|$ is the distance between the sites located at $x$ and $y$, $K_\nu$ is the modified Bessel function of the second kind of order $\nu$ \cite{lebedev2012special}, and we have defined
	\[
	\bar{\lambda}:=\sqrt{\frac{\bar{D}}{\mu}} \qquad \bar{\rho}:=\sqrt{\frac{\sigma^2}{\bar{b}_0}}\quad,
	\]
	where $ \bar{D}:=Da^2 $ and $ \bar{b}_0:=b_0/a^d $ for $ a\to 0$, and they are assumed to be finite in the limit. The first one is a standard scaling for spatial diffusivity, whereas the second scaling assumption comes from the requirement that spatial continuous constants are finite and non-trivial as $a \to 0$ for $x \ne y$.
	
	As the asymptotic behavior of $ K_{\nu} $  as $z \to \infty$ is $K_{\nu}(z)\sim e^{-z} \sqrt{\frac{\pi}{2z}}$, $\bar{\lambda}$ is the correlation length of the system; $\bar{\rho}^2$ has the dimensions of a $ d $-dim volume and gives the local intensity of the correlations. Because eq.(\ref{pp_correlation}) is the continuum limit of eq.(\ref{eq:corr_discrete}), this expression of the pair correlation function is a good approximation of the one in eq.(\ref{eq:corr_discrete}) only when $|x-y|\gg a$ and $ \bar{\lambda}\gg a $.
	
\section{\label{gen-fun} Generating function close to the critical point}

In this section we introduce the parameters which allow us to identify a region close to the critical point of the model. This suggests an expansion which will lead to simplified equations which, nonetheless, carry a lot of information about the model.

A simple way to make progress with the master equation in eq.(\ref{ME}) is the use of a formal Kramers-Moyal expansion \cite{gardiner2004handbook}. It is well-known that there are limitations to this procedure and it has been criticized, because one cannot always pinpoint a small parameter for the correct expansion \cite{gardiner2004handbook,van1992stochastic}. The system-size expansion solves these difficulties, but it has to be applied when the size of the system becomes large \cite{van1992stochastic}. Here, however, it is not entirely evident what parameter should identify the size (population size or volume) of the system in the critical regime. Indeed, the model has no carrying capacity or maximum population size, and the total volume of the system could only provide us with the macroscopic equation, which has no interest in the present case.

In order to make analytical progress, we have introduced two dimensionless parameters, $ \varepsilon $ and $\eta$, which identify a non-trivial region when $ 0<b<r $, but $ b\to r $. This choice comes from the observation that communities of living organisms often appear to have very large demographic fluctuations. Several studies (see \cite{Volkov2005,Volkov2007,tovo2017upscaling}) have pointed out that per capita birth and death rates are close to each other in seemingly different systems, thus suggesting that a sensible theoretical limit to study is when $ b $ approaches $ r $. For example, a fit of eq.(\ref{pp_correlation}) to the empirical two-point correlation function of the tropical forest inventory of Pasoh natural reserve in Malaysia leads to empirical values of $\frac{2(r-b)}{r+b}$ of the order of $10^{-7}$ (see Fig. (\ref{Pasoh_plot}) and section \ref{Eco} for more details). We hence define $\varepsilon:=\frac{2(r-b)}{r+b}$ with the condition that $\frac{b_0}{\mu} \varepsilon =\mathcal{O}(1)$ as $\varepsilon \to 0^+$; in this way we obtain a constant $ \rho^2= \mu/b_0\varepsilon  $, which in real systems is large because usually $ r/b_0\gg 1 $. The parameter $ \varepsilon $ indicates how close the system is to the critical point, regardless of spatial diffusion. With the independent parameter $\eta:=\frac{D}{\sigma^2}$, instead, we compare the importance of spatial diffusion with respect to demographic fluctuations. We will assume that $\eta =\mathcal{O}(\varepsilon)$ as $ \varepsilon\to 0^+ $ and hence $ \eta/\varepsilon=\lambda^2 $, a new independent constant. This scaling assumption reflects that we want to explicitly analyze spatial effects in the critical limit. In fact, it can be easily proved that when $\eta/\varepsilon \to 0$, the model is equivalent to the mean field model without spatial diffusion, whereas for $\eta/\varepsilon \to \infty$ spatial diffusion dominates over the birth-death dynamics.  \\
When $ b $ and $ r $ are close to each other, we expect that population sizes can be well approximated with continuous random variables in each site. In order to understand when this is possible in relation to the parameters $ \varepsilon $ and $ \eta $, we assumed that the generating function $ \zeta(\{H\},t) $ is analytic at $ H_i=0 $ for any $ i $ and the most important contribution to the equation of $ \zeta(\{H\},t) $ comes from a negative neighborhood of the origin with thickness $ \mathcal{O}(\varepsilon) $.  This is tantamount to introducing the change of variable $ H_i=\varepsilon S_i $ into eq.(\ref{eqzeta_full}) and to expanding in powers of $ \varepsilon $, assuming $ S_i=\mathcal{O}(1) $ and $ S_i\leq 0 $. Up to linear order in $ \eta $ and $ \varepsilon $ we obtain
	
	\begin{align}
	\nonumber
	\frac{\partial}{\partial t} \zeta(\{S\},t)=\sum_{i\in \mathbb{L}} \sigma^2 S_i \Big\{ & \eta \Delta_i \frac{\partial \zeta}{\partial S_i}- \varepsilon\frac{\partial \zeta}{\partial S_i}+ \frac{\varepsilon}{\rho^2}\ \zeta +\\
	\label{eqzeta_rescaled3}
	&+\varepsilon S_i \frac{\partial \zeta}{\partial S_i} \Big\} 
	\end{align}
or
\begin{align}
	\nonumber
	\frac{\partial}{\partial T} \zeta(\{S\},t)=\sum_{i\in \mathbb{L}} S_i \Big\{ & \lambda^2 \Delta_i \frac{\partial \zeta}{\partial S_i}- \frac{\partial \zeta}{\partial S_i}+ \frac{1}{\rho^2}\ \zeta +\\
	\label{eqzeta_rescaled4}
	&+ S_i \frac{\partial \zeta}{\partial S_i} \Big\} \quad,
\end{align}
where we have introduced the dimensionless time $T:=\mu t$ and now $ \zeta $, with a slight abuse of notation, indicates the generating function corresponding to continuous (and dimensionless) random variables. Therefore, the evolution equation for the generating function of the population sizes becomes
	\begin{equation}\label{eqzeta_rescaled2}
	\frac{\partial \zeta}{\partial T}=\sum_{i\in \mathbb{L}} H_i \Big\{\lambda^2 \Delta_i \frac{\partial \zeta}{\partial H_i}- \frac{\partial \zeta}{\partial H_i}+ \langle n\rangle\zeta+ \frac{\sigma^2}{\mu}H_i \frac{\partial \zeta}{\partial H_i} \Big\} \quad,
	\end{equation}
	where $ H_i $ is the variable conjugated to the continuous random variable $ n_i $. The parameters also correspond to this continuum limit and now the population sizes $ n_i $ have an exponential cut-off with a (large) characteristic scale given by $ \sigma^2/\mu $. Eq.(\ref{eqzeta_rescaled2}) leads to the following Fokker-Planck equation:
	\begin{align}\label{FP}
	\nonumber
	\frac{\partial}{\partial T} P(\{ n\},T)=&\\
	\nonumber
	\sum_{i \in \mathbb{L}} \Big\{ - \frac{\partial }{\partial n_i} & \Big[\Big( \lambda^2 \Delta_i n_i-n_i+\langle n\rangle \Big) P(\{n\},T) \Big]+\\
	+&\frac{\sigma^2}{\mu} \frac{\partial^2}{\partial n_i^2} \Big[n_i P(\{n\},T) \Big]  \Big\}\quad, \\ \nonumber
	\end{align}
	where now $ \{ n\} $ are continuous random variables and $ P(\{ n\},T) $ is the corresponding probability density function. It is interesting to note that this is not exactly equivalent to the na\"{\i}ve Kramers-Moyal expansion of eq.(\ref{ME}), which would entail additional terms in the diffusive part. Nevertheless it is a diffusive approximation of the process in the regime identified by the parameters $ \eta $ and $ \varepsilon $.

\subsection{Conditional generating function}
Eq.(\ref{FP}) cannot be solved in full generality, but we can better understand the underlying dynamics by studying the distribution of the population sizes in arbitrary volumes of space.
Let us indicate with $\mathcal{V}$ the set of sites belonging to a $ d $-dim volume $|\mathcal{V}|=V$, and let us introduce the random variable $N(V,t)=\sum_{i \in \mathcal{V}} n_i(t)$, i.e. the total number of individuals in $\mathcal{V}$ at time $ t $. By indicating with $P(N|V,t)$ the corresponding probability density function of $N$, we define the conditional generating function
\[
Z(h|V,t)= \langle e^{h N(V,t)} \rangle=\int_{0}^{\infty} \d N \ e^{h N}  P(N|V,t)
\]
where $h\le 0$. We obtain the corresponding equation for $Z$ by specifying $ H_i $, i.e.
\begin{equation}
\label{H_conditions}
H_i=
\begin{cases}
h \text{ if } i \in \mathcal{V}\\
0 \text{ if } i \notin \mathcal{V}\\
\end{cases}
\end{equation}
and substituting this into eq.(\ref{eqzeta_rescaled2}). Thus,
\begin{align}
\label{eqz_rescaled}
\frac{\partial}{\partial T} Z(h|V,T)=h \Big[\lambda^2 &  \sum_{i \in \mathcal{V}} \Delta_i \ f(i,h,V,T)-\frac{\partial Z}{\partial h}+\\
\nonumber
+&\langle n\rangle V Z \Big]+ \frac{\sigma^2}{\mu} h^2 \frac{\partial Z}{\partial h}
\end{align}
where $f(i,h,V,T)=\langle n_i(T) e^{h N(V,T)} \rangle$ and we have used the identity

\begin{equation}\label{identity:f}
\sum_{i \in \mathcal{V}} \langle n_i e^{h N} \rangle=\frac{\partial Z}{\partial h} (h|V,T)\quad.
\end{equation}

This equation depends on $ f(i,h,V,T) $ which in general is unknown. An equation for $ f $ can be derived by differentiating both sides of eq.(\ref{eqzeta_rescaled2}) with respect to $H_k$. Assuming as before that $k \in \mathcal{V}$, we finally obtain the following equation for $f(k,h,V,T)$ (abbreviated $f(k,h)$)
\begin{align}
\label{eqf_inside}
\frac{\partial}{\partial T} f(k,h)=& \lambda^2 \Delta_k f(k,h)- f(k,h)+\\ 
\nonumber
+&\frac{b_0}{\mu} Z+ h \Big[ \lambda^2 \sum_{i \in \mathcal{V}} \Delta_i \ g(i,k,h)+\\ 
\nonumber
-&\frac{\partial }{\partial h} f(k,h)+\langle n\rangle V\, f(k,h) \Big]+\\
\nonumber
+&2 \frac{\sigma^2}{\mu} h f(k,h)+\frac{\sigma^2}{\mu} h^2 \frac{\partial}{\partial h}f(k,h)
\end{align}
where $g(i,k,h,V,T):=\langle n_i(T) n_k(T) e^{h N(V,T)} \rangle$ and where we have used the identity

\begin{equation}\label{identity2}
\sum_{i \in \mathcal{V}} \langle n_i n_k e^{h N} \rangle =\frac{\partial}{\partial h} f(k,h,V,T)\quad.
\end{equation}

In eq.(\ref{eqf_inside}) the function $ g $ is still unknown, but it is possible to show that $\Delta_i g(i,k,h,T) = \Delta_k g(i,k,h,T)$ to leading order as $a \to 0$ (see Appendix \ref{App_B}). This allows us to obtain a closed equation for $ f $. Indeed, we have 
\[
\sum_{i \in \mathcal{V}} \Delta_{i} g(i,k,h,V,T)= \frac{\partial}{\partial h} \Delta_k f(k,h,V,T)
\]
after using the identity (\ref{identity2}).

The spatial continuous limit of eq.(\ref{eqz_rescaled}) as $a \to 0$ reads
\begin{align}
\label{eqz_continuous}
\frac{\partial}{\partial T} Z(h|V,T)=h \Big[\bar{\lambda}^2 &  \int_{\mathcal{V}} \d x \ \nabla_x^2 \ f(x,h,T)-\frac{\partial Z}{\partial h}+\\
\nonumber
+& \langle n \rangle V Z \Big]+ \frac{\sigma^2}{\mu} h^2 \frac{\partial Z}{\partial h}\quad,
\end{align}
where in $ \langle n\rangle $ we have substituted $ b_0 $ with $ \bar{b}_0 $. This equation is of crucial importance in what follows. Similarly, eq.(\ref{eqf_inside}) becomes
\begin{align}
\label{eqf_continuous}
\frac{\partial}{\partial T} f(y,h,V,T)=& \bar{\lambda}^2 \nabla^2_y \ f(y,h)- f(y,h)+\langle n \rangle Z+\\
\nonumber
+& h \Big[ \bar{\lambda}^2 \frac{\partial}{\partial h} \nabla^2_y f(y,h)-\frac{\partial }{\partial h} f(y,h)+\\ 
\nonumber
+& \langle n \rangle V f(y,h) \Big]+\frac{\sigma^2}{\mu} \frac{\partial}{\partial h} \Big[h^2  f(y,h) \Big]\;,
\end{align}
where with the continuous coordinates ($n_i \to n(x)$ and $n_k \to n(y)$), we have also $f(x,h,V,T)=\langle n(x) e^{h N} \rangle$ and $N(V,t)=\int_\mathcal{V} \d x \ n(x,t)$. When there are no spatial effects, i.e., $ \bar{\lambda}=0 $, eq.(\ref{eqz_continuous}) reads
\begin{equation}
	\label{eqz_continuous_mf}
	\frac{\partial}{\partial T} Z(h|V,T)=h \Big[-\left(1- \frac{\sigma^2}{\mu} h \right)\frac{\partial Z}{\partial h}+\langle n \rangle V Z\Big]\quad,
\end{equation}
which has the following solution at stationarity 

\begin{equation}\label{eqz_continuous_mf_stat}
Z(h|V)=\left(1-\frac{\sigma^2}{\mu} h\right)^{-\frac{\mu \langle n \rangle V}{\sigma^2}}\quad,
\end{equation}
where $ h\leq 0 $ and $ \mu \langle n \rangle V/\sigma^2=V/\bar{\rho}^2 $. This is the generating function of a gamma distribution \cite{van1992stochastic} with mean $ \langle n \rangle V $ and variance $ \langle n \rangle^2 V \bar{\rho}^2 $. Also, it is not difficult to verify that when $ \bar{\lambda}=0 $, the solution of eq.(\ref{eqf_continuous}) is given by

\begin{equation}\label{key}
f(y,h,V,T)=\frac{1}{V}\frac{\partial Z}{\partial h}(h|V,T)\quad.
\end{equation}

In the next section we calculate the stationary variance of the population sizes in a volume $ V $, i.e., of the random variable $ N(V) $.

\begin{figure} [ht]
	\centering
	\begin{subfigure}{.49\columnwidth}
		\centering
		\caption{}
		\includegraphics[width=\textwidth]{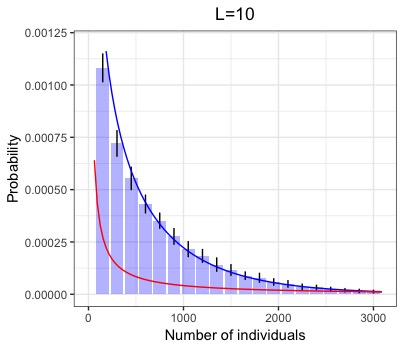}
	\end{subfigure}%
	\hfill
	\begin{subfigure}{.49\columnwidth}
		\centering
		\caption{}
		\includegraphics[width=\textwidth]{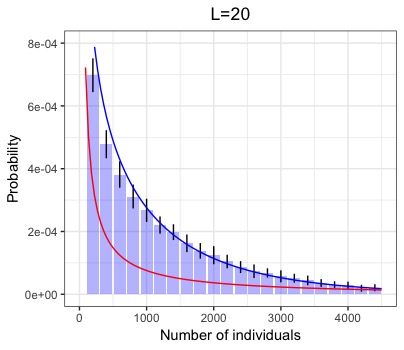}
	\end{subfigure}\\
	\begin{subfigure}{.49\columnwidth}
		\centering
		\caption{}
		\includegraphics[width=\textwidth]{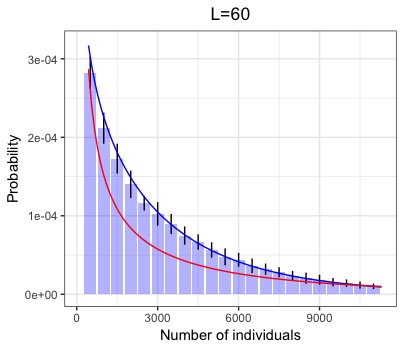}
	\end{subfigure}%
	\hfill
	\begin{subfigure}{.49\columnwidth}
		\centering
		\caption{\qquad }
		\includegraphics[width=\textwidth]{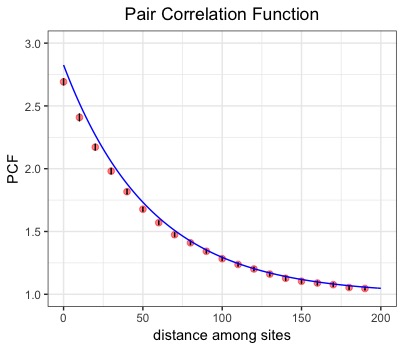}
	\end{subfigure}
	\caption{\textbf{Stationary distribution as obtained from the phenomenological algorithm ($ d=1 $).} Panels (a), (b) and (c) present a comparison between the simulated model as obtained from the phenomenological algorithm (histograms) outlined in sec.\ref{diffreg} and the stationary solution as calculated with eq.(\ref{spatial_stat_sol}) (blue solid line). The lattice comprises $500$ sites in total and we carried out 50,000 independent realizations in $ d=1 $ with periodic boundary conditions, where parameters are $D=30$, $b_0=0.5$, $\mu=0.01$ and $\sigma=10$, and hence $\bar{\lambda} \approx 55$. Panels show results for segments of different lengths which include 10, 20 and 60 adjacent sites, respectively. The size of error bars (black lines) are twice as much the standard deviation, while the red solid line represents the mean field solution of the system (i.e. eq.(\ref{spatial_stat_sol}) where $\Sigma=\sigma^2/\mu$). 
		Panel (d) presents the comparison between the simulated and the analytic pair correlation function at stationarity. Red dots are from simulations, while the blue solid line is the analytic solution given by eq.(\ref{pp_correlation}).}
	\label{Phenom_1d}
\end{figure}

\begin{figure} [ht]
	\centering
	\begin{subfigure}{.49\columnwidth}
		\centering
		\caption{}
		\includegraphics[width=\textwidth]{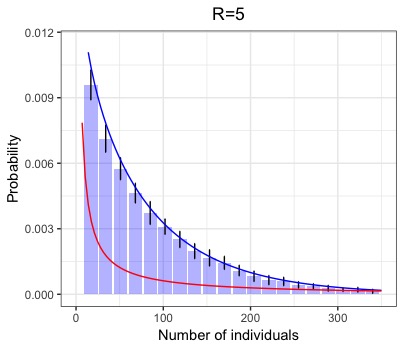}
	\end{subfigure}%
	\hfill
	\begin{subfigure}{.49\columnwidth}
		\centering
		\caption{}
		\includegraphics[width=\textwidth]{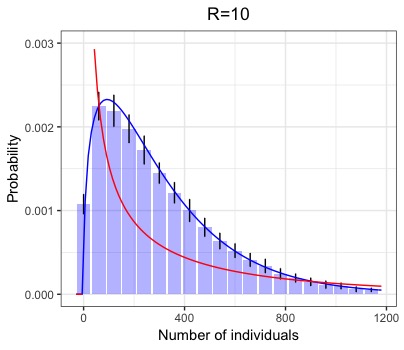}
	\end{subfigure}\\
	\begin{subfigure}{.49\columnwidth}
		\centering
		\caption{}
		\includegraphics[width=\textwidth]{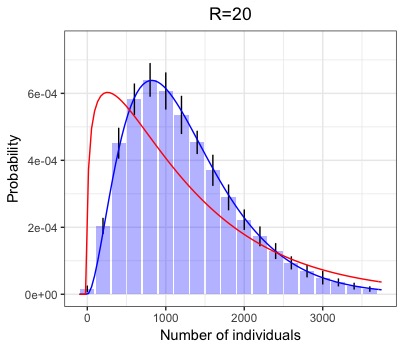}
	\end{subfigure}%
	\hfill
	\begin{subfigure}{.49\columnwidth}
		\centering
		\caption{\qquad }
		\includegraphics[width=\textwidth]{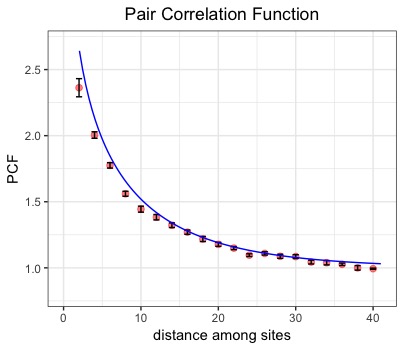}
	\end{subfigure}
	\caption{\textbf{Stationary distribution as obtained from the phenomenological algorithm ($ d=2 $).}  Panels (a), (b) and (c) present a comparison between the simulated model as obtained from the phenomenological algorithm (histograms) outlined in sec.\ref{diffreg} and the stationary solution as calculated with eq.(\ref{spatial_stat_sol}) (blue solid line). The square lattice comprises $200 \times 200$ sites. We carried out 50,000 independent realizations in $ d=2 $ with periodic boundary conditions, with parameters $D=20$, $b_0=0.1$, $\mu=0.1$ and $\sigma=10$, and hence $\bar{\lambda} \approx 15$. Panels (a-c) show results for different areas with radii of length 5, 10 and 20, respectively. The size of error bars (black lines) are twice as much the standard deviation, while the red solid line represents the mean field solution of the system (i.e., eq.(\ref{spatial_stat_sol}) where $\Sigma=\sigma^2/\mu$). Panel (d) presents the comparison between the simulated and the analytic pair correlation function at stationarity. Red dots are from simulations, while the blue solid line is the analytic solution given by eq.(\ref{pp_correlation}).}
	\label{Pheno_2d}
\end{figure}

\section{\label{} Population Variance}

In this section we calculate the stationary variance of the population size, i.e., $ N(V)=\int_\mathcal{V} \d x\, n(x) $, where $ V $ is a finite volume. This quantity is of pivotal importance in the approximation that we are going to develop in the following sections and it is key when introducing spatial information in the equations. Thus we outline here the main calculations, leaving to Appendix \ref{App_C} further details.

We start from the spatial continuous limit of eq.(\ref{pp_equation}) as $a \to 0$, which at stationarity reads

\begin{align}
\nonumber
\bar{\lambda}^2 \Big( \nabla^2_x \langle n(x) n(y) \rangle &+ \nabla^2_y \langle n(x) n(y) \rangle \Big)-2 \langle n(x)n(y) \rangle+ \\ 
\label{pp_equation2}
 & +2 \langle n \rangle^2+2 \frac{\sigma^2}{\mu} \langle n \rangle \delta(x-y)=0
\end{align}
where $\delta(x-y)$ is a Dirac delta and we have included only the leading terms as $ \varepsilon\to 0 $. By integrating both sides of eq.(\ref{pp_equation2}) with respect to $y \in \mathcal{V}$, we obtain an equation for $\langle n(x) N(V) \rangle$. We plan to find an explicit expression for this quantity, because it plays a key role in the following when we will calculate  $ Z(h|V) $. Henceforth, we will take $\mathcal{V}$ to be a $d$-dim ball of radius $R$ and we will assume that the origin of the Cartesian coordinates is at its center. Thus we indicate with $|x|$ the distance from the origin of the site located at $x$ in this coordinate system. With this notation and using the symmetry of $ \langle n(x) n(y) \rangle $ with respect to $ x $ and $ y $, we obtain at stationarity 
\begin{align}
\label{eqboth}
\bar{\lambda}^2 \nabla^2_x\langle n(x) N(R) \rangle-&\langle n(x) N(R) \rangle+\langle n \rangle^2 V+\\
\nonumber
+&\frac{\sigma^2}{\mu} \langle n \rangle \ \Theta \Big( R- |x|  \Big) =0\quad.
\end{align}
This linear ODE has to be solved separately for $|x|>R$ and $|x|<R$. The continuity of $\langle n(x) N(R) \rangle$ and its first derivative at the boundary $|x|=R$ provide the solution. The final results is (see Appendix \ref{App_C})
\begin{align}
\label{sV_correlation}
\langle n(x) N(R) \rangle=&\langle n \rangle^2 V+ \frac{\sigma^2}{\mu} \langle n \rangle \ \Psi \Big(\frac{|x|}{\bar{\lambda}}, \frac{R}{\bar{\lambda}} \Big)\quad,
\end{align}
where the function $\Psi$ takes the following form for $ |x|\leq R $
\begin{align} \label{Psi}
\Psi\Big( \frac{|x|}{\bar{\lambda}} & , \frac{R}{\bar{\lambda}} \Big)=\\
\nonumber
=& 1- \frac{ \Big(\frac{\mid x \mid}{R} \Big)^{1-\frac{d}{2}} \ K_{\frac{d}{2}} \Big( \frac{R}{\bar{\lambda}} \Big) I_{\frac{d}{2}-1} \Big( \frac{\mid x \mid}{\bar{\lambda}} \Big) }{I_{\frac{d}{2}-1} \Big( \frac{R}{\bar{\lambda}} \Big) K_{\frac{d}{2}} \Big( \frac{R}{\bar{\lambda}} \Big)+ I_{\frac{d}{2}} \Big( \frac{R}{\bar{\lambda}} \Big) K_{\frac{d}{2}-1} \Big( \frac{R}{\bar{\lambda}} \Big)}\quad,
\end{align}
where $I_\nu(z)$ and $K_\nu(z)$ are the modified Bessel functions of the first and second kind, respectively \cite{lebedev2012special}. Integrating both sides of eq.(\ref{sV_correlation}) with respect to $x \in \mathcal{V}$, we obtain the equation for the second moment of $N(R)$, i.e.
\begin{align}
\label{NN}
\langle N(R)^2 \rangle=&\langle n \rangle^2 V^2+ \frac{\sigma^2}{\mu} \langle n \rangle V \ \psi \Big( \frac{R}{\bar{\lambda}} \Big)
\end{align}
where $\psi ( R/\bar{\lambda} )$ takes the following explicit form in dimension $d$ (see Appendix \ref{App_C} for its behavior):

\begin{align}
\label{psi}
\psi \Big( \frac{R}{\bar{\lambda}} \Big)= 1- \frac{ \frac{d \bar{\lambda}}{R} \ K_{\frac{d}{2}} \Big( \frac{R}{\bar{\lambda}} \Big) I_{\frac{d}{2}} \Big( \frac{R}{\bar{\lambda}} \Big) }{I_{\frac{d}{2}-1} \Big( \frac{R}{\bar{\lambda}} \Big) K_{\frac{d}{2}} \Big( \frac{R}{\bar{\lambda}} \Big)+ I_{\frac{d}{2}} \Big( \frac{R}{\bar{\lambda}} \Big) K_{\frac{d}{2}-1} \Big( \frac{R}{\bar{\lambda}} \Big)} \quad .
\end{align}
These two functions, namely $\langle n(x) N(R) \rangle$ in eq.(\ref{sV_correlation}) and $\langle N(R) \rangle^2$ in eq.(\ref{NN}) will be used in the next sections to calculate a first order approximation of the spatially explicit probability density function of $N(V)$ in the vicinity of the critical point. 

Finally, these solutions allow us to write down the analytic form of the variance of $N(R)$ in dimension $d$, i.e.
\begin{equation}
\label{variance}
\text{Var}[N(R)]=\langle N(R) \rangle \Sigma(R)\quad,
\end{equation}
where $ \langle N(R)\rangle= \langle n \rangle V =\bar{b}_0V/\mu$ and $\Sigma(R):=\sigma^2 \psi(R/\bar{\lambda})/\mu$. The function $ \Sigma(R) $ is the spatial Fano factor and quantifies the deviations of the fluctuations from a Poisson process. Since $ \sigma^2/\mu=\mathcal{O}((r-b)^{-1}) $, when the system is close to the critical point -- for fixed $ R $ and $\bar{\lambda} $ -- the system has large fluctuations on all scales larger than the correlation length $ \bar{\lambda} $. Also, in the regime $ R/\bar{\lambda}\to +\infty$ we obtain $  \mu \Sigma(R)/\sigma^2 = 1 + \mathcal{O}[(R/\bar{\lambda})^{-1}] $, thus recovering the mean-field fluctuations as predicted by eq.(\ref{eqz_continuous_mf_stat}). 

\subsection{\label{taylor} The spatial Taylor's law}
Taylor's law was first observed in ecological communities \cite{taylor1961aggregation,taylor1977aggregation}, where natural populations show some degree of spatial aggregation. This was phenomenologically captured by assuming a scaling relationship between the variance and mean of population sizes in different areas. More generally, and recently, Taylor's law denotes any power relation between the variance and the mean
of random variables in complex systems \cite{eisler2008fluctuation,james2018zipf}. The law postulates a relation of the following form

\[
\text{Var}[N(R)] = C \langle N(R) \rangle^{\alpha}\quad,
\]
where $ C $ is a positive constant and $\alpha$ typically assumes values between one and two \cite{taylor1977aggregation,james2018zipf}. The spatial model which we have introduced can predict the behavior of this relation across scales without making specific assumptions. If we focus on the two dimensional case and fix $ \bar{\lambda} $, we obtain $\text{Var}[N(R)] = C_1 \langle N(R) \rangle^2 \log ( \langle N(R) \rangle )$ for $R\ll \bar{\lambda}$, while $ \text{Var}[N(R)] =C_2 \langle N(R) \rangle$ for $R \gg \bar{\lambda}$. This latter situation corresponds to the mean field case in which $ C_2=\sigma^2/\mu $. So for areas of radius much smaller than the correlation length the model is characterized by $\alpha=2$ with logarithmic corrections, while in the case of radii much larger than $\bar{\lambda}$ we obtain $\alpha=1$. This is in agreement with previous studies \cite{taylor1961aggregation,taylor1977aggregation,eisler2008fluctuation}. The model predicts $\alpha=1$ at large scales regardless of the dimension of the system, whereas for small areas $ \alpha $ strongly depends on $ d $, e.g. $\alpha=2$ (without logarithmic corrections) in the one dimensional case, and $\alpha=5/3$ for $ d=3 $. Therefore, from small to large spatial scales the model predicts a cross-over of exponents which is difficult to explain without a spatially-explicit framework. Recently, considerable attention has been devoted to the origin of curvatures in scaling relationships, such as those relating body size and the metabolic rates of living organisms (e.g species of mammals \cite{r01748} or freshwater phytoplankton \cite{Zaoli17323}). Unlike what has been found in these latter works, in our model the cross-over in the scaling exponent $\alpha$ is a result of the interplay between space and dispersal at different spatial scales. When considering small areas, local communities appear strongly correlated with each other, while at larger areas these communities are completely independent as they are located at distances much larger than the correlation length. When they are decoupled, the relation is simply $ \text{Var}[N(R)] =b \sigma^2 V(R)/\mu^2$, where $ V(R) $ is the volume of a $ d $-dimensional sphere. 

It has been proved that Taylor's law can emerge in a much more general class of stochastic processes, for example when the dynamical rates of the model are affected by environmental variability \cite{eisler2008fluctuation,james2018zipf}. Here we do not consider this effect, but it would be certainly interesting to investigate environmental stochasticity within a spatially explicit framework.

\section{\label{diffreg} Solution of the conditional pdf}

The goal of this section is to derive the main result of the paper. With the spatial population variance obtained in the previous section along with an appropriate approximation, we are now able to close and solve eq.(\ref{eqz_continuous}) for the conditional generating function $Z(h|V)$. We will then invert this latter for the conditional probability density function $ P(N|V) $, from which several characteristics of the spatial patterns can be deduced.

In order to calculate the explicit form of $Z(h|V,T)$ from eq.(\ref{eqz_continuous}), we first need to calculate the solution of eq.(\ref{eqf_continuous}), which in turn has to satisfy the identity in eq.(\ref{identity:f}). We will first make use of this latter in the form

\begin{equation}\label{identity:f-cont}
\int_{\mathcal{\mathcal{V}}}f(x,h,V,T)\d x=\frac{\partial Z}{\partial h}(h|V,T) \quad.
\end{equation}

It is not difficult to verify that $ f $ can be expressed as

\begin{align}\label{ansatzf1}
f(x,h,V,T)=\frac{1}{V} \frac{\partial Z}{\partial h}\Big[ 1+ \sum_{i=1}^{\infty}h^i A_i(x,V,T) \Big]\quad,
\end{align}
where the functions $ A_i $ are such that

\begin{equation}\label{}
\int_{\mathcal{\mathcal{V}}} A_i(x,V,T)\d x =0
\end{equation}
for any $ i $. One could calculate the explicit form of $ A_i $, which depends on the spatial position, by substituting the expression in eq.(\ref{ansatzf1}) into eq.(\ref{eqf_continuous}). However, the meaning of these functions provides a more efficient way for the calculation. Because of the definition of $ f $, we readily obtain $ f(x,h=0)= \langle n \rangle$ and

\begin{equation}\label{}
\left.\frac{\partial^m f}{\partial h^m}(x,h)\right|_{h=0}= \langle n(x) N^m \rangle
\end{equation}
for $ m=1,2,\ldots $, from which we can make explicit the expression of $ A_i$ by using eq.(\ref{ansatzf1}). For instance, it is not difficult to show (see Appendix \ref{App_E}) that at stationarity

\begin{align*}\label{}
A_1(x)=& \frac{1}{\langle n \rangle} \Big( \langle n(x) N \rangle-\frac{1}{V}\langle N^2 \rangle \Big)=\\
=& \frac{\sigma^2}{\mu} \Big[ \Psi\Big( \frac{\mid x \mid}{\bar{\lambda}}, \frac{R}{\bar{\lambda}} \Big) - \psi\Big( \frac{R}{\bar{\lambda}} \Big) \Big]
\end{align*}
where $ \Psi $ and $ \psi $ have been defined in eqs.(\ref{Psi}) and (\ref{psi}), respectively. Similar relations, though more complicated, hold for $ i=2,3,\ldots $. Actually, $ \langle n(x) N^i \rangle $ can be calculated by integrating $ i $ times over the volume $ \mathcal{V} $ the $ (i+1) $-th spatial correlation function. Nonetheless, note that this result depends on the symmetry of the $ d $-dim volume $\mathcal{V}$, and on its connectedness. This is important for exploiting the spherical symmetry of the system when introducing polar coordinates, and in selecting its center as the origin (see also the previous section and Appendix \ref{App_C}) .

Since we are interested in relatively large population sizes when the correlation length is either large or small ($\bar{\lambda} \to 0,\infty$ but finite), we retain only the first two terms in the bracket of eq.(\ref{ansatzf1}). Thus, $ f $ at stationarity turns into 

\begin{equation}\label{ansatzf-final}
f(x,h,V)=\frac{1}{V} \frac{\partial Z}{\partial h}\Big[1+ h A_1(x)\Big]\quad,
\end{equation} 
where $ A_1 $ is the one we have obtained before. It is remarkable that, when substituting eq.(\ref{ansatzf-final}) into eq.(\ref{eqz_continuous}), at stationarity one obtains (see Appendix \ref{App_E})

\begin{align}
\label{eqz_ansatz}
\Big(1-h \Sigma(R) \Big) \frac{\partial Z}{\partial h}=\langle n \rangle V Z \quad,
\end{align}
where $\Sigma(R)=\sigma^2 \psi(R/\bar{\lambda})/\mu$ is the spatial Fano factor defined in the previous section. The form of eq.(\ref{eqz_ansatz}) and that of  eq.(\ref{eqz_continuous_mf}) at stationarity are the same, provided we replace $ \sigma^2/\mu $ with $ \Sigma(R)$. Therefore the solution of eq.(\ref{eqz_ansatz}) is

\begin{equation}\label{eqz_continuous_space_stat}
Z(h|R)=\Big( 1-\Sigma(R) h \Big)^{-\frac{\langle n \rangle V(R)}{\Sigma(R)}}\quad,
\end{equation}
which, when inverted for the probability density function, gives a gamma distribution of the form
\begin{align}
	\label{spatial_stat_sol}
	P(N|R)=\Big( \frac{1}{\Sigma(R)} \Big)^{\frac{\langle n \rangle V(R)}{\Sigma(R)}} & \frac{N^{\frac{\langle n \rangle V(R)}{\Sigma(R)}-1} e^{-\frac{N}{\Sigma(R)}}}{\Gamma\left(\frac{\langle n \rangle V(R)}{\Sigma(R)}\right)}\quad.
\end{align}
Thus, while without spatial effects the characteristic scale of the population size is $ \epsilon^{-1} \simeq \sigma^2/\mu \gg 1$, in this regime space introduces a space-dependent scale for fluctuations which is quantified by $ \Sigma(R)=\epsilon^{-1}\psi(R/\bar{\lambda})$, where $ \psi $ is the function defined in eq.(\ref{psi}).

As a further insight, if one defines the new process for the random variable $ N(R,T) $ in the volume $ V $ of fixed radius $ R $ by the stochastic differential equation

\begin{equation}\label{eq:implicit-space}
\dot{N}(R) = \bar{b}_0 V(R)-\mu N(R) + \sigma \sqrt{\psi(R/\bar{\lambda})N(R)}\xi(t)\quad,
\end{equation}
where $ \xi(t)$ is a zero-mean Gaussian white noise and $ \langle \xi(t)\xi(t')\rangle=2\delta(t-t') $, then the stationary pdf of $ N(R) $ is exactly eq.(\ref{spatial_stat_sol}). Notice that space is taken into account only implicitly through the functions $ V(R) $ and $ \psi(R/\bar{\lambda}) $. Eq.(\ref{eq:implicit-space}) can also be obtained as an $ \epsilon $-limit of a spatially-implicit master equation along the lines we have showed in Section \ref{gen-fun}. This process -- unlike the spatial one -- satisfies the detailed balance condition at stationarity as the flux at $ N=0 $ is set to zero. This result suggests that there are some families of spatially-explicit processes which, when restricted to a finite volume, can be well approximated by spatially-implicit processes. Whilst the former brakes detailed balance, the latter turns out to be simpler and satisfies the detailed balance condition. In this model the region of this approximation is close to the critical point of the process. 

It remains to understand when the $ f $ in the form of eq.(\ref{ansatzf-final}) solves eq.(\ref{eqf_continuous}), i.e. the original equation for $ f $. In Appendix \ref{App_E} we show that, for a fixed radius $ R $, at stationarity $ f $ is a solution when $ h $ is small and $ \bar{\lambda} $ is either very large or very small (but finite) compared to $ R $, regardless of the spatial dimension $ d $. The limits $ \bar{\lambda}\to 0 $ and $ \bar{\lambda}\to +\infty $ of eq.(\ref{spatial_stat_sol}) lead to the mean-field expressions, respectively eq.(\ref{eqz_continuous_mf}) and $ P(N|R)=\delta(N-\langle n \rangle V(R)) $ at leading order. Thus, eq.(\ref{spatial_stat_sol}) captures the leading behavior of the distribution of the random variable $ N(R) $ in the large population regime and in the vicinity of the critical point. The simulations indeed confirm this with very good accuracy as shown in Figs.(\ref{Phenom_1d}) and (\ref{Pheno_2d}).

Because we assumed that $ A_1(x) $ is at stationarity, eq.(\ref{ansatzf-final}) does not give all the correct terms for a time evolution of the process $ N(R,T) $. However, relatively close to stationarity, even the temporal dynamics is accurately described by eq.(\ref{eq:implicit-space}). We have checked this idea and compared the exact simulations of the process -- as provided by the Doob-Gillespie algorithm (see also the following section) -- with the corresponding analytic temporal evolution as obtained from solving eq.(\ref{eq:implicit-space}). Assuming that initially in the volume $ V$ there are $ N_0 $ individuals, we find the following time-dependent solution (see \cite{azaele2006dynamical} for the details of the derivation):
\begin{widetext}
\begin{align}
\label{spatial_time_sol}
P(N,t|N_0,0)=\Big( \frac{1}{\Sigma(R)} \Big)^{\frac{\bar{b}_0 V}{\mu \Sigma(R)}} & N^{\frac{\bar{b}_0 V}{\mu \Sigma(R)}-1} e^{-\frac{N}{\Sigma(R)}} \ \frac{\Big[\Big(\frac{1}{\Sigma(R)} \Big)^2 \ N_0 N \ e^{-\mu t} \Big]^{\frac{1}{2}-\frac{\bar{b}_0 V}{2 \mu \Sigma(R)}}}{1-e^{- \mu t}} \\
\nonumber
&\exp\Big[ -\frac{\frac{1}{\Sigma(R)}(N+N_0) e^{-\mu t}}{1-e^{-\mu t}} \Big] I_{\frac{\bar{b}_0 V}{\mu \Sigma(R)}-1} \Big[ \frac{\frac{2 }{\Sigma(R)}\sqrt{N_0 N e^{- \mu t}}}{1-e^{- \mu t}} \Big]\quad,
\end{align}
\end{widetext}
where we used reflecting boundary conditions at $ N=0 $ at any $ t>0 $. This pdf indeed tends to the stationary solution in eq.(\ref{spatial_stat_sol}) as $t \to \infty$. The agreement between simulations and eq.(\ref{spatial_time_sol}) is showed in Fig.(\ref{time_MF_compare}) and (\ref{time_3curve}).\\

\subsection*{Simulations with an efficient algorithm}

These families of spatial stochastic models are difficult to simulate with parameters in arbitrary regimes and even so, usually only on relatively small lattices in low dimensions. Indeed, it is difficult to assess whether or not simulations have reached stationarity (because of the effect of large fluctuations) and how many replicates are necessary to get reliable predictions for the spatial moments. In this light, it is even more important to know the analytical behavior of some quantities, which could not have been guessed from the simulations only. We have therefore compared our analytic predictions to simulations as obtained from a range of different parameter sets and from two different simulation schemes.

The first one is the Doob-Gillespie's algorithm \cite{Gillespie1977exact} for producing exact trajectories of Markovian processes. We have used periodic boundary conditions in 1-$ d $ lattices of various sizes. Parameters were chosen so that the correlation length of the system was much smaller than the total size of the lattice. We have analyzed different sets of parameters and for each one we have run 50,000 independent realizations: error bars were calculated by grouping the results into 50 sets made of 1000 realizations each. In principle this simulation scheme allows us to obtain the exact trajectories of the system at any time, from an initial configuration up to stationarity. However, it is computationally very expensive, and therefore we have been forced to choose relatively small lattice sizes to investigate significant changes from the initial configuration. The results of this are shown in Figs.(\ref{time_MF_compare}) and (\ref{time_3curve}).

In order to analyze a wider set of parameters and larger lattices, we have simulated the process by using a new algorithm which generates the stationary random field obtained from the stochastic partial differential equation defined on the lattice. This was introduced in \cite{peruzzo2017phenomenological}, and modifies a previous scheme that was used for simulating models of directed percolation \cite{Dornic2005integration}. Here we briefly summarize the main steps of the pseudo-code. From the definition of the discrete Laplace operator, we split the term $\lambda^2 \Delta_i n_i$ in eq.(\ref{FP}) into one part depending only on $n_i$ ($2^d \ \lambda^2 n_i$) and another one depending on the densities in the nearest neighboring sites ($\lambda^2 \sum_{j:|i-j|=1} n_j$). Conditional on the values of $n_j$ for $j \ne i$, the second term is constant and thus $P(n_i|n_j)$ can be obtained as a gamma distribution at stationarity \cite{peruzzo2017phenomenological}. Starting from a random initial configuration, at each iteration $m$, we randomly select a site $i$ and update the value of $n^m_i$ in the next step by sampling from $n^{m+1}_i \sim P(n_i|n^m_j)$, conditional on the values of $n^m_j$ (i.e. $n^{m+1}_j=n^{m}_j$). These steps are repeated until lattice configurations become independent of initial conditions. While standard stochastic integration schemes fail to preserve the positivity of $\{n\}$ at any step, this algorithm produces non-negative populations by construction. We have verified that the simulated distribution thus obtained for $N(R)$ matches the exact simulations of the Doob-Gillespie's algorithm in $d=1 $ and for different lengths, when the comparison was feasible (see Fig.(\ref{Phenom_results})). By means of this algorithm we were able to study much larger lattice sizes at stationarity, and compare simulations against the predictions of the analytic solutions. The agreement was excellent in all the expected regimes (see Figs.(\ref{Phenom_1d}) and (\ref{Pheno_2d})).

\section{\label{Eco} An ecological application}

\begin{figure} [ht]
	\centering
	\begin{subfigure}{.49\columnwidth}
		\centering
		\caption{}
		\includegraphics[width=\textwidth]{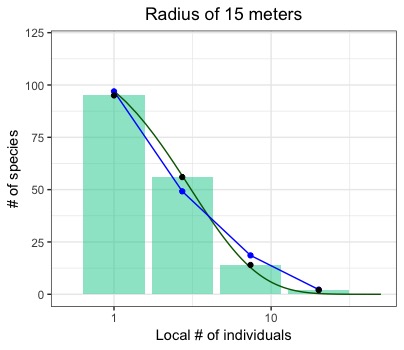}
	\end{subfigure}%
	\hfill
	\begin{subfigure}{.49\columnwidth}
		\centering
		\caption{}
		\includegraphics[width=\textwidth]{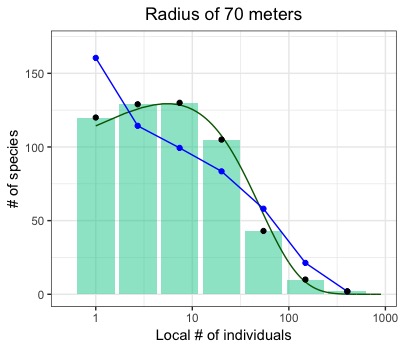}
	\end{subfigure}\\
	\begin{subfigure}{.49\columnwidth}
		\centering
		\caption{}
		\includegraphics[width=\textwidth]{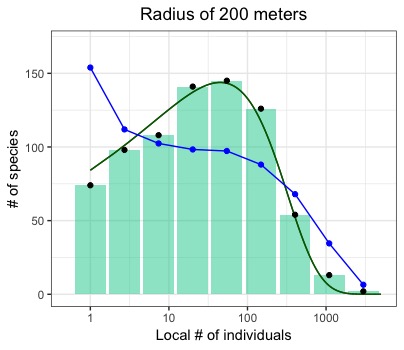}
	\end{subfigure}%
	\hfill
	\begin{subfigure}{.49\columnwidth}
		\centering
		\caption{\qquad }
		\includegraphics[width=\textwidth]{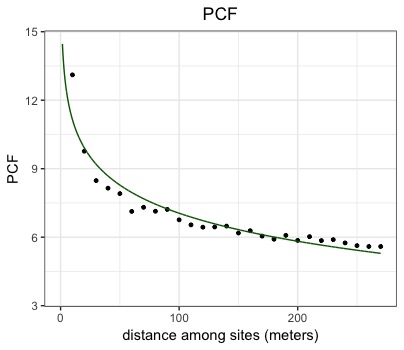}
	\end{subfigure}
	\caption{\textbf{Species abundance distribution from a lowland tropical forest and prediction from the model.} Panels (a), (b) and (c) present the comparison between the empirical data of the distribution of  species' abundances from the lowland tropical forest inventory of the Pasoh natural reserve (Malaysia) and the prediction based on the model. Histograms and  black points represent the empirical data, the green solid lines are the predictions obtained from the solution of the model, i.e. from eq.(\ref{spatial_stat_sol}), and the blue solid lines are those from the best-fit of the Fisher log-series. We highlight that the green lines are not best fits to the empirical data, but genuine predictions that are formulated from the empirical measures of the pair correlation function, as described in the main text (see eq.(\ref{pp_correlation})). The radii of the areas are $15, 70$ and $200$ meters as reported on the corresponding panels. For the statistical analysis see Appendix \ref{AppE}. In panel (d) we compare the empirical pair correlation function (black dots) with the best fit of eq.(\ref{pp_correlation}). The correlation length that is thus calculated is $\bar{\lambda} \approx 2.5 \times 10^3$ meters, with $\bar{\rho} \approx 8.9 \times 10^3$ and $\langle n \rangle \approx 6.1 \times 10^{-4}$ trees per square meter for each species. The total number of species in the whole 50Ha forest stand is $\approx 900$. }
	\label{Pasoh_plot}
\end{figure}

A simple, but far from trivial, application of the mathematical model we have previously described is the modelling of spatial patterns in ecosystems with a large number of species. Examples include the species richness and abundance distribution of coral reefs, bees across landscapes or vascular plant species in tropical forest inventories. Starting with the crude approximation that species are independent (at least at some relatively large scale), this model can be used to predict the abundance distribution of species from measures of the two-point correlation function (PCF) and mean abundance per species. Because these latter descriptors are relatively easy to calculate, it turns out that we can obtain an estimate of how many rare (or abundant) species live in a region without surveying the entire system. Therefore, as well as being theoretically interesting, this approach has also an important practical advantage, because it ultimately allows one to infer the total number of species within a very large spatial region by utilizing only scattered and small-scale samples of the region itself. This is a long-lasting problem which has received a lot of attention recently \cite{tovo2017upscaling,shem2017solution,azaele2016statistical}.

Our goal here is not to explain the upscaling method, but only show to what degree the spatial model is in agreement with ecological empirical data. The two-point correlation function (PCF) specifies how similar individuals are distributed as a function of the geographic distance. Since species are assumed to be independent, the spatial distribution of individuals belonging to the same species can be considered as an independent realization of the stochastic process. As a consequence, the term $\langle n(x) n(y) \rangle/\langle n \rangle^2$ in eq.(\ref{pp_correlation}) can be easily calculated as the product of local abundances of individuals in sites that lay at the same distance, averaged across all species.  Usually, it is more likely that close-by individuals belong to the same species than individuals that live farther apart. This translates into a PCF that is always positive, but decays with distance \cite{condit2002beta, chave2002spatially}. A decline in similarity with increasing geographic distance indicates that the individuals of a community are spatially aggregated. Therefore, a simple random placement of individuals in space is not a good approximation of the configurations of the community as thought in the past \cite{coleman1981random}. On the contrary, the stationarity PCF of this model decays with distance, is always positive (see eq.(\ref{pp_correlation})) and has two free parameters ($ \bar{\lambda} $ and $ \bar{\rho} $). When we calculate these latter from a best fit of the empirical data, we obtain a good agreement as shown in Fig.(\ref{Pasoh_plot}d).

The mean abundance per species is readily available, because the total number of species and individuals are known in this forest plot. Since at stationarity the model is fully specified by these three parameters ($ \langle n \rangle $, $ \bar{\lambda} $ and $ \bar{\rho} $), we are henceforth able to predict all the stationary patterns that we like to compare with those of the empirical ecosystem. One of them is the probability that a species has a given number of individuals within a specific region. In the ecological literature it is often referred to as species abundance distribution (SAD) \cite{mcgill2007species}. In our model it is given by eq.(\ref{spatial_stat_sol}) and the histograms of this pattern are reported in Fig.(\ref{Pasoh_plot}a-c) for different radii. The SAD represents one of the most commonly used static measures for summarizing information on ecosystem's diversity. Interestingly, the shape of the SAD in tropical forests has been observed to maintain similar features, regardless of the geographical location or the details of species interactions. Indeed, it often displays a unimodal shape at larger scales and a peak at small abundances at relatively small spatial scales. Numerous papers have focused on evaluating the processes that generate and maintain such observed characteristics \cite{black2012stochastic, azaele2016statistical}, but only few of them considered spatial effects in an explicit framework \cite{pigolotti2018stochastic, o2018cross}.

The panels in Fig.\ref{Pasoh_plot} show a comparison between the empirical data from the forest inventory of Pasoh Natural reserve in Malaysia (year 2005) and the predicted abundance distribution of species obtained from the model we have described in the previous sections. It is remarkable that the predicted curves in the first three panels are genuine inferences obtained from the mean abundance per species (i.e., $ \langle n \rangle $) and the PCF (i.e., $ \bar{\lambda} $ and $ \bar{\rho} $), and not best fitted curves to empirical SAD. As a comparison, we have included the best fit to data of a probability distribution commonly used as null model in the literature, the Fisher log-series (see Appendix \ref{AppE}) \cite{azaele2016statistical}. The two methods have comparable accuracies at smaller scales, while at larger scales our method outperforms the Fisher log-series and captures the empirical SAD with much higher accuracy.

Such an agreement confirms that abundances of species are indeed characterized by very large fluctuations and that local populations appear correlated over very large spatial scales. This entails that complex ecosystems may comprise a large number of rare species, whereas only a few have large abundances (hyperdominant species). Because this model does not explicitly include interactions nor environmental forcing, we cannot single out which ecological processes bring about this finding. Nevertheless, the model is able to explain this separation of population size scales by poising a system close to criticality.
 
In terms of new physical insight, the theoretical framework that we have presented here allows us to connect exactly the two-point correlation function to the Species Abundance Distribution (SAD) and the Species Area Relationship (SAR) of a system. This explains quantitatively why and how SAR (known as $ \alpha $-diversity in the ecological literature) and species spatial turnover (known as $ \beta $-diversity in the ecological literature) are related. For instance, it is usually assumed that the SAR is a power law function of the area \cite{hubbell2001unified}, i.e. $ S(A)=cA^z $, however our results show that this can only be an approximation which works on a given range of spatial scales.

The simplicity and generality of the model make it suitable for the description of patterns in other biological systems. Indeed, numerous recent studies have showed that spatial bio-geographical patterns emerge in marine ecosystems \cite{rinaldo2002cross, ser2018ubiquitous}, microorganisms \cite{woodcock2007neutral}, including bacteria \cite{giometto2014emerging, giometto2015sample}, arch\ae{}a, viruses, fungi \cite{hanson2012beyond} and eukaryotes \cite{altermatt2015big,lynch2015ecology}. Future work will include the application of the current framework to those biological communities.

\section{Conclusions}

In this paper we have studied a spatial stochastic model which can be fruitfully used to describe the main large scale characteristics of species-rich ecosystems. We have shown how to calculate analytically some of the most important spatial patterns when the system is close to criticality, which is the regime where the most important features emerge.

The model encapsulates birth, death, immigration and local hopping of individuals. It describes the dynamics of point-like and well-mixed individuals living in a metacommunity defined on a $d$-dimensional regular graph. The model is also minimal, meaning that, without one of its components (i.e., birth, death, nearest-neighbor hopping and external immigration), it yields either trivial or well-known results. Despite its simplicity, however, it violates detailed balance (see Appendix A) and generates a remarkable phenomenology of patterns, when the birth and death rates are comparable (criticality), leading to its properties being governed by large fluctuations, whose effect of course escape any classical mean-field analysis. These patterns and fluctuations entail strong correlations on large spatial and temporal scales. 

The linearity of the rates is not sufficient to derive a full solution (in a weak sense) of the model. However, in applications one is usually interested in the analytical properties of processes that are much less general than the spatial random field. Thus we restricted our analysis to the conditional distribution, $ p(N|V) $, that $ N $ individuals are found in a volume $ V $. This quantity is sufficiently general to describe a wealth of patterns in several systems. We have found that in the close-to-critical regime, $ p(N|V) $ satisfies an aptly derived equation which has the form of the corresponding mean-field equation of the process (i.e., without space). This equation includes functions of $ V $, which we have exactly calculated. Such spatial redefinition of the parameters introduces strong deviations from the corresponding mean-field solutions, as confirmed by the exact stochastic simulations. Also, this shows that the process that governs the random variable $ N $ satisfies detailed balance in a first approximation and close to the critical point, thus being considerably simpler than the distribution of the random field. This result suggests that not only it is possible to make considerable analytical progress in this model, but there may be other, more general, models close to criticality which can be studied with a similar approach. 

Indeed, our results suggest the tantalizing hypothesis that the conditional distribution, $ p(N|V) $, provided by some families of spatial stochastic processes, is described by much simpler processes within a specific region of the parameter space. In our model the region of this approximation is close to the critical point of the process and the distribution of the simpler process holds the same functional shape across all spatial scales. Of course, the hypothesis requires much more scrutiny, especially when spatial models include nonlinear terms which can jeopardize the methods developed here. On the other side, our approach allows for the analysis of several generalizations, including non-local dispersal kernels and different sources of noise (e.g., environmental noise).

On a more applied side, our framework explains how the most important patterns in macro-ecology (e.g., SAR, SAD and PCF) are intrinsically connected with each other (see eqs.(\ref{pp_correlation}) and (\ref{spatial_stat_sol}) and their relation through $ \Sigma(R) $ in eq.(\ref{variance})), and also accounts for a cross-over of power law behaviors in the spatial Taylor's law (see eq.(\ref{variance})). We have compared the predictions against those of the Fisher log-series, commonly used as “null model” in the ecological literature. Despite the free parameter of the Fisher log-series distribution was calculated directly from the empirical data of the species abundance distribution (best fit) at each spatial scale, our model performed much better (see Fig.4), even though we employed the parameters obtained from the PCF curve, which does not provide information about the distribution of abundances of species. Discrepancies between empirical data and theoretical predictions will potentially inform us on the importance of alternative physical effects which will be included in more realistic models. 

We have finally shown that, when applying the model to large ecosystems, species are predicted to display a broad range of abundances as a consequence of the critical regime. Therefore, many of them are rare and only a few are very abundant. These large demographic fluctuations are correlated across large spatial scales as well as over long times, in agreement with several empirical datasets collected in well-known tropical forest inventories \cite{peruzzo2017phenomenological,tovo2017upscaling,azaele2015towards}.  

Of course, these findings do not prove that these biological systems are close to criticality, but suggest that it is worth pursuing that route further. In this article we have not investigated the reasons why those systems look nearly critical, nor how they can operate within such tiny regions of their parameter space. For this, one needs to look into how inter- and intra-interactions dynamically lead living systems towards the correct region, in which states are biologically meaningful -- however, see \cite{mastromatteo2011criticality, goudarzi2012emergent, stieg2012emergent, hidalgo2014information}. Nonetheless, our results could help understand what key factors drive such dynamics, and possibly shed light on the importance and effects of non-linearities among interacting agents.

\section{Acknowledgments}

FP thanks the NERC SPHERES DTP (NE/L002574/1) for funding his studentship. We thank Amos Maritan for insightful discussions.

\begin{appendices}
	
\section{\label{BDB} Broken Detailed Balance}
We here recall the main general properties of detailed balance. Let us denote with $\textbf{c}$ the configuration of a generic stochastic process and indicate with $p(\textbf{c},t|\textbf{c}_0,t_0)$ the probability that the configuration $\textbf{c}$ is seen at time $t$, given that the configuration at time $t_0$ was $\textbf{c}_0$ (abbreviated $p(\textbf{c},t)$). We introduce $\mathcal{W}(\textbf{c}'|\textbf{c})$, which is the (time-independent) rate to transit from state $\textbf{c}$ to $\textbf{c}'$. If we consider Markovian dynamics, the evolution of $p(\textbf{c},t)$ is given by the following master equation (ME) \cite{van1992stochastic,gardiner2004handbook}
\begin{equation} \label{ME_generic}
\frac{\partial p(\textbf{c},t)}{\partial t}= \sum_{\textbf{c}'} \Big[ \mathcal{W}(\textbf{c}|\textbf{c}') p(\textbf{c}',t)-\mathcal{W}(\textbf{c}'|\textbf{c}) p(\textbf{c},t) \Big] \quad.
\end{equation}
If it happens that $ \mathcal{W}(\textbf{c}|\textbf{c}') P(\textbf{c}')-\mathcal{W}(\textbf{c}'|\textbf{c}) P(\textbf{c})=0 $ for all configurations (this is the detailed balance (DB) condition), then the probability distribution $ P(\textbf{c}) $ is also a stationary solution of the ME, as we see from eq.(\ref{ME_generic}). On the other hand, it is clear that not all stationary distributions satisfy the detailed balance condition.
\begin{figure}[h]
	\includegraphics[width=\linewidth]{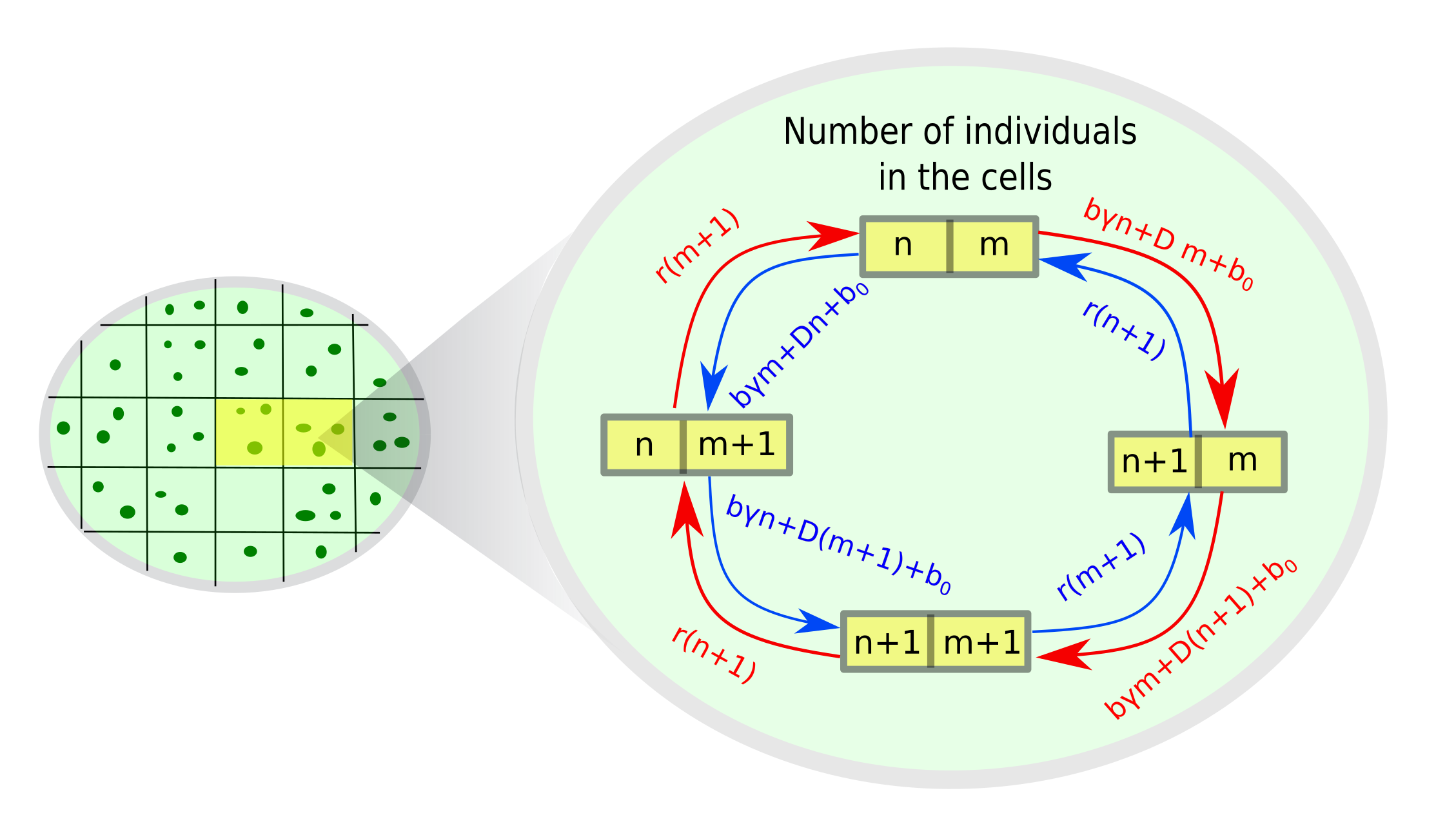}
	\caption{\textbf{Broken Detailed Balance.} The picture represents a closed path in the space of configurations of the process which has different probabilities depending on the direction of the path (see also \cite{grilli2012absence}). We take two neighboring sites, which initially have $n$ and $m$ individuals. The rates of jumping into the next configuration of the path are reported in the image and the arrows indicate the direction. In this simple case $ D=b(1-\gamma) $. In the clockwise direction (red arrows) the total rate is $[b \gamma n+Dm+b_0][ b \gamma m +D (n+1)+b_0][r(n+1)][r(m+1)]$. In the anti-clockwise direction (blue arrows) the total rate is $[b \gamma m+Dn+b_0][ b \gamma n +D (m+1)+b_0][r(m+1)][r(n+1)]$. These two total rates must be equal for the detailed balance to hold. However, for any arbitrary configuration ($m \ne n$) this is true only when $D(D-b\gamma)r=0$, i.e. for $b = 0$, $r= 0$, $\gamma = 1, 1/2$.}
	\label{DB_violation}
\end{figure}

It is possible to show \cite{zia2007probability} that a condition for the validity of DB is that the probability of following a closed path in the space of configurations does not depend on the orientation of the path. More precisely, DB is satisfied if and only if for any choice of a closed path $\{\textbf{c}_1,...,\textbf{c}_m\}$, with $m$ an arbitrary number, the following holds
\begin{align} \label{DB_condition}
	\mathcal{W}(\textbf{c}_1|\textbf{c}_2)& \mathcal{W}(\textbf{c}_2|\textbf{c}_3) \cdots \mathcal{W}(\textbf{c}_{m}|\textbf{c}_1)=\\ \nonumber
	=&\mathcal{W}(\textbf{c}_{1}|\textbf{c}_m) \mathcal{W}(\textbf{c}_{m}|\textbf{c}_{m-1}) \cdots \mathcal{W}(\textbf{c}_2|\textbf{c}_1)\; .
\end{align}
This equation corresponds to microscopic reversibility and, when it is violated, the system can be found in non-equilibrium steady states.

Violation of DB is key to living systems and is the subject of intense recent interest \cite{gnesotto2018broken,battle2016broken}. In fig. (\ref{DB_violation}) we show with a simple example that for this model the condition in eq.(\ref{DB_condition}) is not satisfied, hence the detailed balance condition does not hold: Fig. (\ref{DB_violation}) shows a path in the space of configurations for which the total rate does depend on the orientation of the closed path. Notice that such counterexample does not hold when spatial dispersal is switched off (i.e. $D=0$ or $ \gamma=1 $), or when autocatalytic production and spatial dispersal rates are equal ($ \gamma=1/2 $).

\section{\label{App_A} Spatial correlation of the birth-death Markov process}
In the main text we have defined the spatial generating function of the model
\begin{align*}
\zeta(\{H\},t)=&\langle e^{\sum_{k\in \mathbb{L}}n_k H_k} \rangle=\\
=&\sum_{\{n\}} e^{\sum_{k\in \mathbb{L}}n_k H_k} p(\{n\},t)\quad.
\end{align*}
From eq.(\ref{ME}), multiplying both sides through by $e^{\sum_{s\in \mathbb{L}} n_s H_s}$ and summing over all states, we find eq.(\ref{eqzeta_full}) of the main text. If we differentiate by $H_k$ and impose $\{H\}=0$, we find the equation for the mean number of individuals $\langle n_k \rangle$, reported in the main text. Taking another derivative with respect to $H_l$ and setting $\{H\}=0$, we obtain the equation for the spatial two-point correlation function (PCF) among the sites $k$ and $l$
\begin{align*}
\frac{\partial}{\partial t} \langle n_k n_l \rangle=D & \Big( \Delta_l \langle n_k n_l \rangle + \Delta_k \langle n_k n_l \rangle \Big) -2 \mu \langle n_k n_l \rangle+\\ 
+&2 b_0 \ \langle n \rangle+ \delta_{k,l} \Big( 2 \sigma^2 \langle n \rangle +b_0+ D \Delta_k \langle n_k \rangle \Big)\;,
\end{align*}
where $\sigma^2:=\frac{b+r}{2}$, $D:=\frac{b(1-\gamma)}{2 d}$, $\delta_{k,l}$ is a Kronecker delta, and $\Delta$ is the discrete Laplace operator as defined in the main text.\\
Considering stationary patterns, because of homogeneity we have $\Delta_k \langle n_k \rangle=0$ and, introducing $G_{k,l}=\langle n_k n_l \rangle-\langle n \rangle^2$, we obtain
\begin{align*}
D  (\Delta_k G_{k,l}+& \Delta_l G_{k,l})-2 \mu G_{k,l}+\\ 
+&\Big( 2 \sigma^2 \langle n \rangle + b_0 \Big) \delta_{k,l}=0\quad.
\end{align*}

We will now consider the Fourier series expansion of $G_{\textbf{k},\textbf{l}}$, which we will write as
\[
G_{\textbf{k},\textbf{l}}=\Big(\frac{a}{2 \pi} \Big)^d \int_\mathcal{C} \d \textbf{p} \ \hat{G}(\textbf{p}) e^{i \textbf{p} \cdot (\textbf{k}-\textbf{l})}\quad,
\]
where $ \textbf{k},\textbf{l} $ are the Cartesian coordinates of the locations of sites on the lattice (in $ a $ units), $\textbf{p}$ is a vector with $d$ components which belongs to $\mathcal{C}$, the hypercubic $d$-dimensional primitive unit cell of size $2 \pi/a$. Upon substituting $G_{\textbf{k},\textbf{l}}$ in the stationary equation, we get an expression for $\hat{G}(\textbf{p})$, which is 
\[
\hat{G}(\textbf{p})= \Big( \frac{\sigma^2}{\mu} \langle n \rangle + \frac{b_0}{2 \mu} \Big) \frac{1}{1+\frac{2D}{\mu}\sum_{i=1}^{d}(1-\cos(p_i a))} \quad,
\]
where $ p_i $ is the $ i $-th component of $ \textbf{p} $. Therefore 

\begin{align*}
G_{\textbf{k},\textbf{l}}=\Big(\frac{a}{2 \pi} \Big)^d \int_\mathcal{C} \d \textbf{p} \dfrac{\Big( \frac{\sigma^2}{\mu} \langle n \rangle + \frac{b_0}{2 \mu} \Big)e^{i \textbf{p} \cdot (\textbf{k}-\textbf{l})}}{1+\frac{2D}{\mu}\sum_{i=1}^{d}(1-\cos(p_i a))}\quad.
\end{align*}
We can obtain a good deal of simplification by taking a continuous spatial limit ($a \to 0$). Renaming $\textbf{p} \to p$, $\textbf{k} \to x$ and $\textbf{l} \to y$ (now continuous variables in $ \mathbb{R}^d $), and appropriately rescaling the constants as explained in the main text, we arrive at the expression for the pairwise spatial correlation in dim $d$, i.e.
\begin{align*}
G(x,y)= \frac{\bar{\rho}^2 \langle n \rangle^2}{(2 \pi)^d } \Big( 1+\frac{\mu}{2 \sigma^2} \Big) \int \d p \ \frac{e^{i p \cdot (x-y)}}{1+ \bar{\lambda}^2 p^2}\quad,
\end{align*}
where we have replaced $ b_0 $ with $ \bar{b}_0 $ in $ \langle n \rangle $ and used the definitions of $\bar{\rho}, \bar{\lambda} $ given in the main text. Actually, the $ d $-dim integral in the previous expression can be calculated 

\begin{align*}
	\int\!\! \d p \,\frac{e^{i p \cdot (x-y)}}{1+ \bar{\lambda}^2 p^2}&=\int_{0}^{\infty}\!\!\d s \int \!\! \d p\, e^{i p \cdot (x-y)-(1+ \bar{\lambda}^2 p^2)s}\\
	&=\int_{0}^{\infty}\!\!\d s \left(\frac{\pi}{s}\right)^{\frac{d}{2}}e^{-\frac{|x-y|^2}{4\bar{\lambda}^2 s}-s}\\
	=\frac{(2 \pi)^{d/2}}{ \bar{\lambda}^d} & \Big( \frac{\mid x-y \mid}{\bar{\lambda}} \Big)^{\frac{2-d}{2}} K_{\frac{2-d}{2}} \Big(\frac{\mid x-y \mid}{ \bar{\lambda}} \Big)
\end{align*}
where $K_a(z)$ is the modified Bessel function of the second kind and in the last step we used 9.6.24 from \cite{abramowitz1965handbook}. Eventually, the PCF is
\begin{align}
\label{eqcorrMP}
G(x,y)=\frac{\bar{\rho}^2 \langle n \rangle^2}{(2 \pi \bar{\lambda}^2)^{d/2}} & \Big( 1+\frac{\mu}{2 \sigma^2} \Big) \times\\
\nonumber
\times & \Big( \frac{\mid x-y \mid}{\bar{\lambda}} \Big)^{\frac{2-d}{2}} K_{\frac{2-d}{2}} \Big(\frac{\mid x-y \mid}{ \bar{\lambda}} \Big)\quad.
\end{align}
Because $K_a(z)$ decays exponentially for large $ z $ regardless of $ a $, $ \bar{\lambda} $ plays the role of a correlation length of the spatial system. Instead, $ \bar{\rho}^2 $ has dimensions of a $ d $-dim volume and provides a characteristic volume of local demographic fluctuations.

\begin{figure}
	\includegraphics[width=\columnwidth]{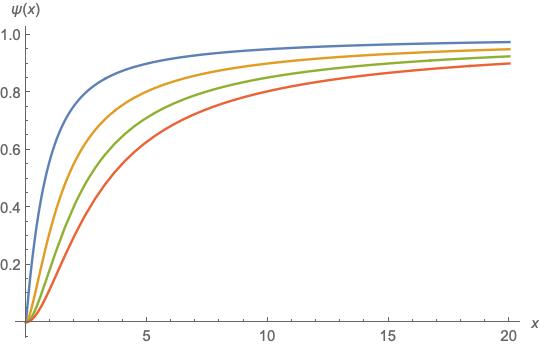}
	\caption{\textbf{Behavior of the function $ \psi(x) $.} The figure shows the behavior of $  \psi(x) $ as defined in eq.(\ref{psi}) in dim $ d=1 $ (blue solid curve), $ d=2 $ (yellow solid curve), $ d=3 $ (green solid curve) and $ d=4 $ (red solid curve).}
	\label{psix}
\end{figure}

\begin{figure} [ht]
	\centering
	\begin{subfigure}{.49\columnwidth}
		\centering
		\caption{}
		\includegraphics[width=\textwidth]{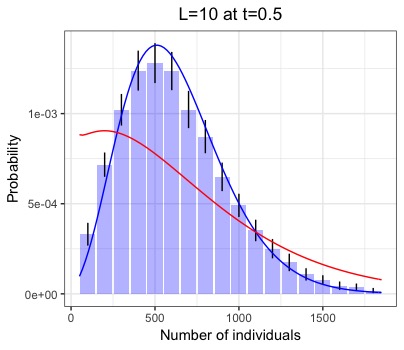}
	\end{subfigure}%
	\hfill
	\begin{subfigure}{.49\columnwidth}
		\centering
		\caption{}
		\includegraphics[width=\textwidth]{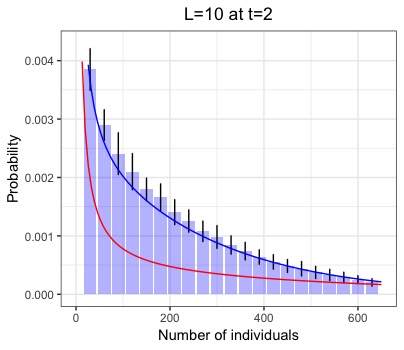}
	\end{subfigure}\\
	\begin{subfigure}{.49\columnwidth}
		\centering
		\caption{}
		\includegraphics[width=\textwidth]{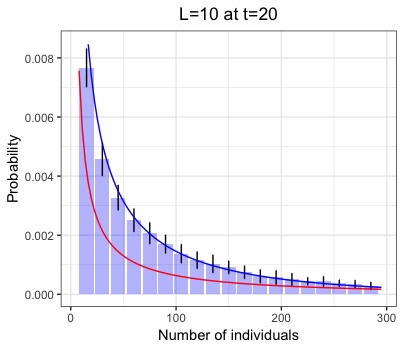}
	\end{subfigure}%
	\hfill
	\begin{subfigure}{.49\columnwidth}
		\centering
		\caption{\qquad }
		\includegraphics[width=\textwidth]{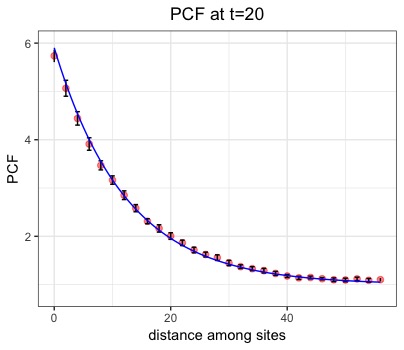}
	\end{subfigure}
	\caption{\textbf{Time evolution with the Doob-Gillespie algorithm.} Panels (a), (b) and (c) present a comparison between the simulated model as obtained from the Doob-Gillespie algorithm (histograms), the analytic prediction as defined in eq.(\ref{spatial_time_sol}) (blue solid line), and the mean field solution (red line). Simulations were carried out in $ d=1 $ with periodic boundary conditions, where each site initially contained exactly $100$ individuals, with a total number of $200$ lattice sites; $ L $ comprises $10$ adjacent sites and time $ t $ is expressed in units of $\mu^{-1}$. The size of error bars are twice as much the standard deviation. Results in panel (d) are from the same set of simulated data (red dots, with error bars) and compares simulations to the analytic curve (blue line) of the pair correlation function at large times ($t=20$). Parameters in these simulations are  $b=600$, $r=601$, with $\gamma=0.5$ and $b_0=5$ ($\lambda \approx 12$).}
	\label{time_MF_compare}
\end{figure}

\begin{figure}[h!]
	\includegraphics[width=\linewidth]{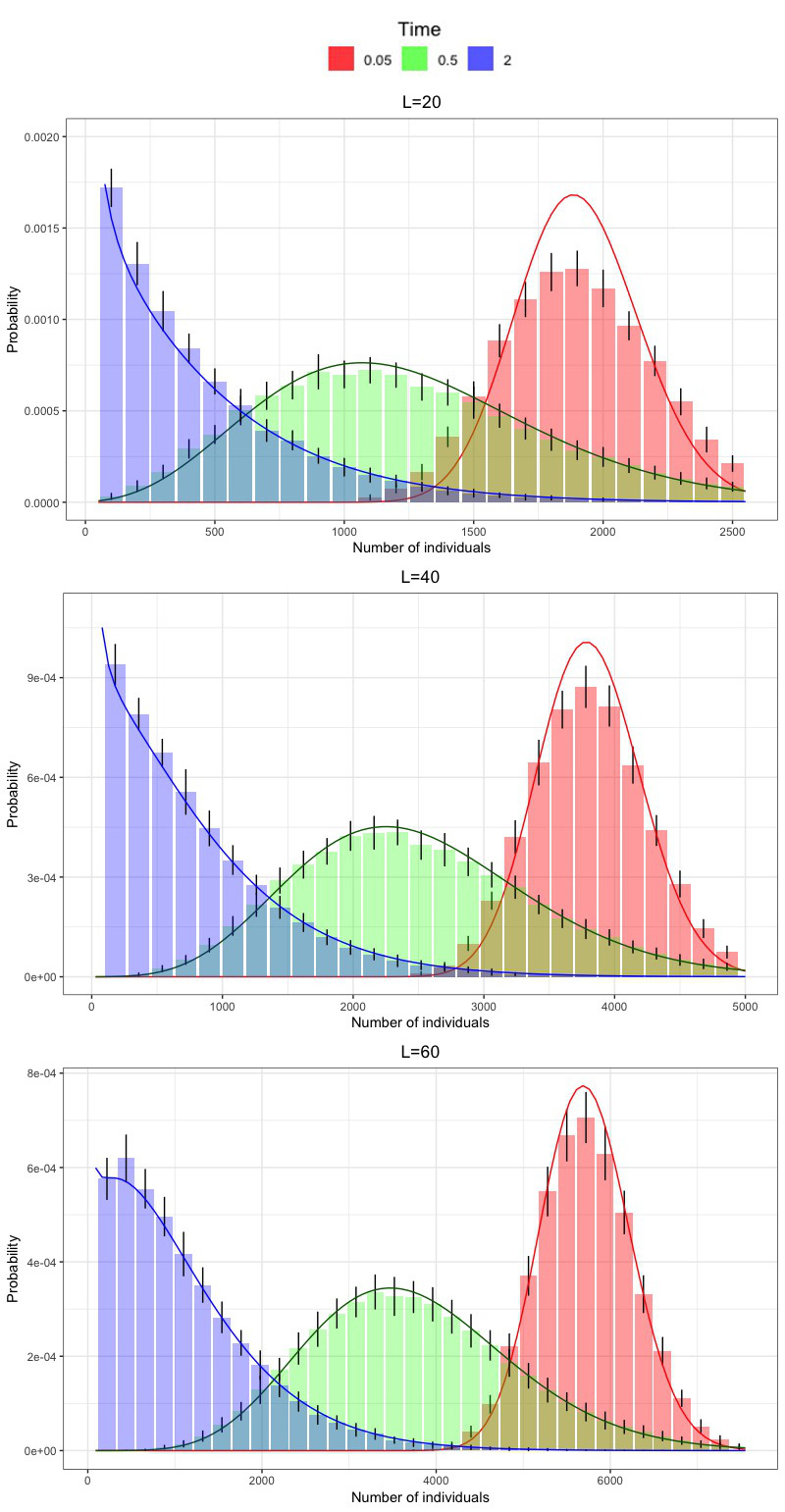}
	\caption{\textbf{Time comparison for different areas and different times in one dimension.} Simulations were carried out in $ d=1 $ with periodic boundary conditions, where each site initially contained exactly $100$ individuals, with a total number of $200$ lattice sites; parameters are the same as in Fig.(\ref{time_MF_compare}). The three panels show comparisons of simulated data from the Doob-Gillespie algorithm and the analytic formula presented in eq.(\ref{spatial_time_sol}), and at different segment lengths ($20$, $40$ and $60$ sites, respectively), and within each panel the three plots refer to different times $T$ ($T=0.05$ for red histograms, $T=0.5$ for green histograms and $T=2$ for blue histograms, units of $ \mu^{-1} $). Histograms represent data from simulations, solid lines are the analytic predictions and error bars have length twice as much the standard deviation. For small $ T $ predictions and simulated data differ, as expected. However, as we approach stationarity the prediction  improves significantly, and already at $T=0.5$ the analytical and simulated distribution match very well.}
	\label{time_3curve}
\end{figure}

\begin{figure}
	\includegraphics[width=\textwidth,height=4cm]{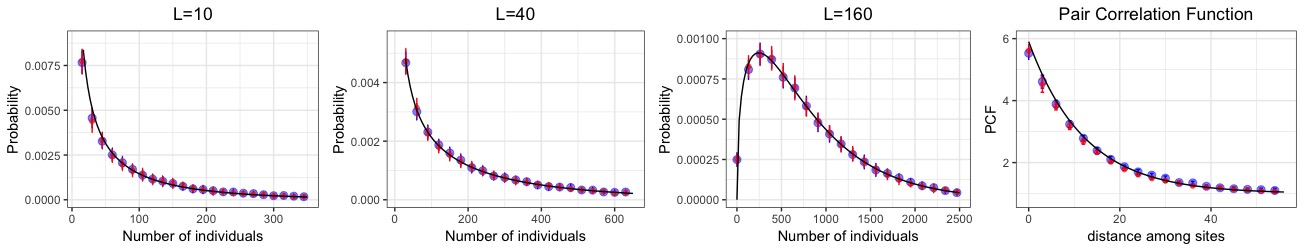}
	\caption{\textbf{Comparison between Gillespie and Phenomenological simulation scheme.} The four panels show the conditional pdf, $ P(N|L) $, simulated with the same set of parameters and on lattices of the same size, but using different algorithms at stationarity and in one dimension. Blue dots represent data from the phenomenological scheme (and blue lines are the respective error bars), while red dots are from Doob-Gillespie algorithm (with respective error bars). Black solid lines are the analytic predictions from eq.(\ref{spatial_stat_sol}). The parameters are $b=600$, $d=601$, $\gamma=0.5$, $b_0=5$ ($D=150$) and hence $\lambda \approx 12$. The lattices have $200$ total sites with periodic boundary conditions. }
	\label{Phenom_results}
\end{figure}

\section{\label{App_B} Approximating $g(i,j,h,t)$ in the continuous spatial limit}

Equation (\ref{eqf_inside}) shows that the evolution of $f(k,h,t)$ is not closed, being coupled to the generating function $Z$ and to the function $g(i,k,h,t)= \langle n_i n_k e^{h N(V)} \rangle$. In this section we want to show that, at leading order in the limit $a \to 0$, the value of $\Delta_i g(i,k,h,t)$ approaches $\Delta_k g(i,k,h,t)$. This makes the equation for $f(k,h,t)$ much simpler, basically decoupling it with the other quantities. For simplicity we will only consider the one dimensional case, but this claim holds true in higher dimensions as well. Thus $ \mathcal{V} $ will be an interval of length $2R$ and the origin of the coordinate system will be located at its center. By changing slightly the notation, we will now indicate $N(V)=N(-R,R)$, thus making explicit that $V$ extends from site $-R$ to $R$. Since the system is spatially homogeneous, we can write
\begin{align*}
\langle n_{i-a} & n_k e^{h N(-R,R)} \rangle=\\
&=\langle n_{i} n_{k+a} e^{h N(-R+a,R+a)} \rangle
\end{align*}
As $a\to 0$ and at leading order, we can neglect $a$ in the argument of $N(-R,R)$, thus obtaining
\begin{align*}
g(i-a,k,h,t)=& \langle n_{i-a} n_k e^{h N(-R,R)} \rangle=\\
=&\langle n_{i} n_{k+a} e^{h N(-R,R)} \rangle=\\
=&g(i,k+a,h,t)
\end{align*}
Similarly $\langle n_{i+a} n_k e^{h N(-R,R)} \rangle=\langle n_{i} n_{k-a} e^{h N(-R,R)} \rangle$. Thus, as $a \to 0$ at leading order we can write
\begin{align*}
\Delta_i g(i,k)=&g(i+a,k)+g(i-a,k)-2 g(i,k)=\\
=& g(i,k-a)+g(i,k+a)-2 g(i,k) \\ 
=&\Delta_k g(i,k)\quad.
\end{align*}
This approximation makes it possible to obtain eq.(\ref{eqf_continuous}) in the continuous limit.

\section{\label{App_C} Multidimensional Variance}
In the main text we have outlined how to calculate the second moment of the random variable $ N(V) $. Here we provide some more details. We will take $\mathcal{V}$ to be a $d$-dimensional ball with the origin of the Cartesian coordinates in its center. By integrating $ \langle n(x)n(y)\rangle $ over $y \in \mathcal{V}$ in eq.(\ref{pp_equation2}) and using the symmetry of  $ \langle n(x)n(y)\rangle $ in $ x $ and $ y $, we obtain an equation for $\langle n(x)N(V)  \rangle$, which at stationarity reads

\begin{align}\label{eq:app_nN}
	\nonumber
	\frac{\bar{D}}{\mu} \nabla^2_x \langle n(x) N(V) \rangle & - \langle n(x) N(V)  \rangle+ \frac{\bar{b}_0}{\mu} \frac{\bar{b}_0 V}{\mu}+\\ 
	+&\frac{ \sigma^2}{\mu} \frac{\bar{b}_0}{\mu} \Theta(R-|x|) =0\quad,
\end{align}
where $\Theta(z)$ is the Heaviside step function and $ V $ is the $ d $-dim volume

\[
V=\frac{\pi^{d/2}}{\Gamma\Big(\frac{d}{2}+1 \Big)} \ R^d
\]
with radius $ R $. Eq.(\ref{eq:app_nN}) must be solved for $R<|x|$ and $R>|x|$ separately. Boundary conditions for $\langle n(x) N(V) \rangle$ and its first derivative will fix the values of the integrating constants. For $|x|<R$ we obtain
\begin{align*}
\langle n(x) N(V) \rangle=\Big( \frac{\bar{b}_0}{\mu} \Big) \frac{\bar{b}_0 V}{\mu}+ \frac{ \sigma^2}{\mu} \frac{\bar{b}_0}{\mu}+\\
+A \Big( \frac{|x|}{\bar{\lambda}} \Big)^{1-\frac{d}{2}} & I_{\frac{d}{2}-1} \Big( \frac{|x|}{\bar{\lambda}} \Big)
\end{align*}
while for $|x|>R$ instead
\begin{align*}
\langle n(x) N(V) \rangle&=B \Big( \frac{|x|}{\bar{\lambda}} \Big)^{1-\frac{d}{2}}K_{\frac{d}{2}-1} \Big( \frac{|x|}{\bar{\lambda}} \Big)+\Big( \frac{\bar{b}_0}{\mu} \Big) \frac{\bar{b}_0 V}{\mu}\quad.
\end{align*}
The constants $A$ and $B$ will be fixed using the aforementioned continuity conditions. Upon explicit calculation, for $|x| \leq R$ we obtain
\begin{widetext}
\begin{align}
\label{pa_correlation}
\langle n(x) N(V) \rangle=\Big( \frac{\bar{b}_0}{\mu} \Big)^2 V+&\frac{ \sigma^2}{\mu} \frac{\bar{b}_0}{\mu} \Big[ 1-\Big( \frac{|x|}{R} \Big)^{1-\frac{d}{2}} \frac{K_{\frac{d}{2}} \Big( \frac{R}{\bar{\lambda}} \Big)I_{\frac{d}{2}-1} \Big( \frac{|x|}{\bar{\lambda}} \Big)}{I_{\frac{d}{2}-1} \Big( \frac{R}{\bar{\lambda}} \Big) K_{\frac{d}{2}} \Big( \frac{R}{\bar{\lambda}} \Big)+ I_{\frac{d}{2}} \Big( \frac{R}{\bar{\lambda}} \Big) K_{\frac{d}{2}-1} \Big( \frac{R}{\bar{\lambda}} \Big)}\Big]\nonumber\\
=\Big( \frac{\bar{b}_0}{\mu} \Big)^2 V+&\frac{ \sigma^2}{\mu} \frac{\bar{b}_0}{\mu}\Psi\Big( \frac{|x|}{\bar{\lambda}}, \frac{R}{\bar{\lambda}}\Big)\quad,
\end{align}
where $\Psi(a,b)$ was defined in eq.(\ref{Psi}) of the main text and $I_\nu(z)$ and $K_\nu(z)$ are the modified Bessel functions of the first and second kind, respectively \cite{lebedev2012special}. Integrating with respect to $x \in \mathcal{V}$ and using the properties of $I_\nu(z)$, we can readily obtain the explicit form of the second moment of $N(V)$, i.e.
\begin{align}
\label{nn_correlation}
\langle N(V)^2 \rangle=\Big( \frac{\bar{b}_0 V}{\mu}\Big)^2+ \frac{ \sigma^2}{\mu} \frac{\bar{b}_0 V}{\mu} \Big[ 1-& \frac{d \ \bar{\lambda}}{R} \ \frac{K_{\frac{d}{2}} \Big( \frac{R}{\bar{\lambda}} \Big) I_{\frac{d}{2}} \Big( \frac{R}{\bar{\lambda}} \Big) }{I_{\frac{d}{2}-1} \Big( \frac{R}{\bar{\lambda}} \Big) K_{\frac{d}{2}} \Big( \frac{R}{\bar{\lambda}} \Big)+ I_{\frac{d}{2}} \Big( \frac{R}{\bar{\lambda}} \Big) K_{\frac{d}{2}-1} \Big( \frac{R}{\bar{\lambda}} \Big)}\Big]\quad,
\end{align}
where we can read off the explicit expression for $\psi(R/\lambda)$, also reported in eq.(\ref{psi}). Finally, since $\text{Var}(N(V))=\langle N(V)^2 \rangle-\langle N(V) \rangle^2$ we can explicitly write down the variance of $N(V)$, which takes the form
\[
\text{Var}(N(V))= \frac{\bar{b}_0 V}{\mu} \frac{ \sigma^2}{\mu} \Big[ 1- \frac{d \ \bar{\lambda}}{R} \ \frac{K_{\frac{d}{2}} \Big( \frac{R}{\bar{\lambda}} \Big) I_{\frac{d}{2}} \Big( \frac{R}{\bar{\lambda}} \Big) }{I_{\frac{d}{2}-1} \Big( \frac{R}{\bar{\lambda}} \Big) K_{\frac{d}{2}} \Big( \frac{R}{\bar{\lambda}} \Big)+ I_{\frac{d}{2}} \Big( \frac{R}{\bar{\lambda}} \Big) K_{\frac{d}{2}-1} \Big( \frac{R}{\bar{\lambda}} \Big)}\Big]=\langle N(V) \rangle \Sigma(R)\quad,
\]
where $\Sigma(R/\bar{\lambda})=\sigma^2 \psi(R/\bar{\lambda})/\mu$. $ \Sigma(R/\bar{\lambda}) $ is the spatial Fano factor and quantifies the relative importance of fluctuations in the system. Finally, Fig.(\ref{psix}) shows the behavior of $ \psi(x) $.
\end{widetext}

\section{\label{App_E} Evaluation of the regimes of accuracy of the method}

In this section we show that the truncation of $f(x,V,h)$ as in eq.(\ref{ansatzf-final}) yields an accurate approximation for the conditional probability distribution of the model. By retaining the first two terms in the square brackets of eq.(\ref{ansatzf1}), at stationarity we are left with the following 
\begin{align}
\label{ansatzf}
f(x,V,h)=\frac{1}{V} \frac{\partial Z}{\partial h} & \Big[ 1+ h A_1( x ,R) \Big]\quad,
\end{align}
where $Z(h)$ is the conditional generating function at stationarity. By taking the derivative with respect to $ h $ of both sides of eq.(\ref{ansatzf}) and setting $h=0$, we obtain 
\[
\frac{\partial f}{\partial h} \Big|_{h=0}=\langle n(x) N(V) \rangle= \frac{1}{V} \langle N(V)^2 \rangle+\langle n \rangle A_1(x,R)
\]
which gives
\begin{align*}
A_1(x ,V)=\frac{1}{\langle n \rangle} \Big( \langle n(x) N(V) \rangle -\frac{1}{V} \langle N^2 \rangle \Big)\quad.
\end{align*}
Because in Appendix \ref{App_C} we have already calculated $\langle n(x) N(V) \rangle$ and $\langle N(V)^2\rangle$ (see eqs.(\ref{pa_correlation}) and (\ref{nn_correlation})), $ A_1(x ,V) $ is known explicitly. Substituting $ f(x,V,h) $ in eq.(\ref{ansatzf}) with $A_1(x ,V) $ obtained before into eq.(\ref{eqz_continuous}) at stationarity, we get 

\begin{align*}
   \frac{h}{\langle n \rangle V} \frac{\partial Z}{\partial h}\, \bar{\lambda}^2 \!\!\int_\mathcal{V} dx & \ \nabla^2_x \langle n(x) N(V) \rangle -\frac{\partial Z}{\partial h}+\\ 
+&\langle n \rangle V Z+\frac{\sigma^2}{\mu} h \frac{\partial Z}{\partial h}=0\quad.
\end{align*} 
Now we can readily simplify the term $\bar{\lambda}^2 \nabla^2_x \langle n(x) N(V) \rangle$ by making use of eq.(\ref{eq:app_nN}). Integrating this latter with respect to $ x $ in $ \mathcal{V} $, since $ \langle N(R)\rangle=\langle n \rangle V $ and $\Sigma(R)=(\langle N(R)^2 \rangle-\langle N(R) \rangle^2)/\langle N(R)\rangle$ (see Appendix \ref{App_C}), we are therefore left with the following equation
\begin{align}
\label{eqz_ansatz-appendix}
\Big(1-h \ \Sigma(R) \Big)\frac{\partial Z}{\partial h}=\langle n \rangle V  Z \quad,
\end{align}
which therefore provides the equation for the generating function of $ N(R) $ up to terms $ \mathcal{O}(h) $.

Similarly to what we have done so far, we can get further insight into the evolution of $ f(x,V,h) $. We substitute $ f(x,V,h) $ from eq.(\ref{ansatzf}) into eq.(\ref{eqf_continuous}) and use eq.(\ref{eq:app_nN}) to obtain $\bar{\lambda}^2 \nabla^2_x \langle n_x N(V) \rangle$. Eventually, at stationarity this yields
\begin{align}
\label{eqf_ansatz_full}
\frac{\partial}{\partial h} \Big\{ h & \Big[ -\Big(1-h \Sigma(R) \Big) \frac{\partial Z}{\partial h} +\langle n \rangle  V Z \Big] \Big\}+\\
\nonumber
&+ A_1(x,R) \Big\{ \frac{\partial}{\partial h} \Big[ \frac{\sigma^2}{\mu} h^3  \ \frac{\partial Z}{\partial h} \Big]+ h^2 \ \langle n \rangle V \ \frac{\partial Z}{\partial h} \Big\} =0\;.
\end{align}
The first addend of eq.(\ref{eqf_ansatz_full}) is zero because of eq.(\ref{eqz_ansatz-appendix}), and the remaining terms are negligible when $ hA_1$ is small.
 
Fixing the values of $|x|$, $R$ and $\sigma^2/\mu$, we can calculate explicitly the regimes of $\bar{\lambda}$ where $A_1(x,R)$ approaches zero.
From eqs.(\ref{Psi}) and (\ref{psi}) in the main text it is easy to rewrite $A_1(x,R)$ as
\[
A_1(x,R)=\frac{\sigma^2}{\mu} \Big[ \Psi \Big( \frac{|x|}{\bar{\lambda}}, \frac{R}{\bar{\lambda}} \Big)-\psi \Big( \frac{R}{\bar{\lambda}} \Big) \Big]
\]
The asymptotic expansion of the modified Bessel functions is \cite{lebedev2012special}  
\begin{align*}
I_\nu(z)=\frac{e^z}{\sqrt{2 \pi z}}\Big(1+\mathcal{O}[1/z] \Big)\\
K_\nu(z)=e^{-z} \sqrt{\frac{\pi}{2 z}}\Big(1+\mathcal{O}[1/z] \Big)
\end{align*}
when $z\to \infty$. From this it is not difficult to verify that as $\bar{\lambda} \to 0$ we have
\begin{align*}
\Psi \Big( \frac{\mid x \mid}{\bar{\lambda}}, \frac{R}{\bar{\lambda}} \Big)&=1+\mathcal{O}\Big[e^{-\frac{R-\mid x \mid}{\bar{\lambda}}} \Big( \frac{\mid x \mid}{R}\Big)^{1-d}\Big] \\
\psi \Big( \frac{R}{\bar{\lambda}} \Big) &=1+\mathcal{O}\Big( \frac{\bar{\lambda}}{R} \Big)
\end{align*}
and so for $|x|< R$ indeed $A_1(x,R) \to 0$.\\
The case of $\bar{\lambda} \to \infty$ is more elaborate. Let's call $z=R/\bar{\lambda}$. After some lengthy but otherwise straightforward calculations we can verify that as $z\to 0$ we can write
\begin{align*}
\psi(z)=
\begin{cases}
z+\mathcal{O}(z^2) \qquad  &\text{for}\; d=1\\
 -\frac{z^2}{2} \ \log(z)+\mathcal{O}(z^2)   \qquad &\text{for}\; d=2\\
\frac{2 z^2}{5}+\mathcal{O}(z^3) \qquad &\text{for}\; d=3
\end{cases}
\end{align*}
and at leading order $\psi(z)$ is proportional to $z^2$ for $d\geq 3$.\\
For the case of $\Psi$ let us write $z_x=|x|/\bar{\lambda}$. We can verify that
\begin{align*}
\Psi(z_x,z)=
\begin{cases}
\nonumber
z+\mathcal{O}(z^2,z_x^2) \qquad  &\text{for}\; d=1\\
 -\frac{z^2}{2} \ \log(z)+\mathcal{O}(z^3,z_x^2)   \qquad &\text{for}\; d=2\\
 \nonumber
\frac{z^2}{3}+\mathcal{O}(z^3,z_x^2) \qquad &\text{for}\; d=3
\end{cases}
\end{align*}
and at leading order $\Psi$ is proportional to $z^2$ for $d\geq 3$. Thus again $A_1(x,R) \to 0$. These findings follow from that $ f $ goes into mean-field regimes as $\bar{\lambda} \to 0, \infty$ (but finite) and hence all spatial terms go to zero.

\section{\label{AppF} Model with independent dispersal}
In this section we show how the analysis that is undertaken in the main text can be extended also to a model where spatial dispersal is independent of birth.

As in the main text, the new model consists of a spatial meta-community where local communities are located on a $d$-dimensional regular graph (or lattice), and within each community individuals are treated as diluted, well-mixed and point-like particles.

The model is defined by the following birth-death dynamics: each individual dies at a constant death rate $r$ and gives birth at a constant rate $b$, and communities are also colonized from the outside at a constant immigration rate $b_0$ However, individuals can now jump from one site to any of its 2$d$ nearest neighboring sites at any time (not just after birth), and this happens with rate $D$. 

Indicating with $X_i$, $i \in \mathbb{L}$, the individual living in site $i$, the reactions defining the model's dynamics are the following
\begin{align*}
X_i \xrightarrow{b}& \ 2 X_i\\
X_i \xrightarrow{D}& \ X_j\\
X_i \xrightarrow{r}& \ \emptyset\\
\emptyset \xrightarrow{b_0}& \ X_i
\end{align*}
where $j$ indicates a nearest neighbour of site $i$. Comparison with the reactions reported in  section \ref{ME_model_par} shows that, indeed, now spatial movement is decoupled from birth events.

Let us now indicate with $n_i$ the number of individuals in site $i$ and with  $P(\{n\},t)$ the probability to find the system in the configuration $ \{n\} $ at time $ t $. The master equation for $P(\{n\},t)$ then reads

\begin{align}
\nonumber
\frac{\partial}{\partial t}  P(\{n\} & ,t)=\sum_{i \in \mathbb{L}} \Big\{ [b \ (n_i-1)+b_0] P(\{...n_i-1, ...\},t)+\\
\label{ME_movement}
+& r (n_i+1) \ P(\{...n_i+1, ...\},t)+\\
\nonumber
-&r n_i \ P(\{n\},t) - [b n_i+b_0] P(\{n\},t)\\
\nonumber
+&D \sum_{j: |j-i|=1} [(n_j+1) P(\{...n_i-1,n_j+1, ...\},t)]\\ 
\nonumber
-& D \ 2d \ n_i P(\{n \},t) \Big\}
\end{align}
where the dots represent that all other occupation numbers remain as in $\{ n \}$ and it is intended that $P(\cdot)=0$ whenever any of the entrances is negative.

Similarly to the procedure followed in the main text, as a first step we introduce the spatial generating function of the model, defined  in eq.(\ref{gen_func_def}). Multiplying through eq.(\ref{ME_movement}) by $e^{\sum_{k\in \mathbb{L}}n_{k} H_{k}}$ and averaging, we obtain the equation for $\zeta(\{H\},t)$, which reads
\begin{align}
\label{eqzeta_movement}
\frac{\partial}{\partial t} \zeta(\{H\},t)=& \sum_{i \in \mathbb{L}} \Big\{ D \sum_{j: |j-i|=1} \Big[ (e^{H_i-H_j}-1)\frac{\partial \zeta}{\partial H_j} \Big]+\\ 
\nonumber
+&b \ (e^{H_i}-1) \frac{\partial \zeta}{\partial H_i} +b_0 (e^{H_i}-1) \zeta(\{H\},t)+\\
\nonumber
+& r (e^{-H_i}-1) \frac{\partial \zeta}{\partial H_i} \Big\}\quad.
\end{align}
The next steps follow exactly the same procedure undertaken in the main text.

Thus, we define the parameter $\varepsilon=\frac{2(r-b)}{r+b}$ and assume the following parameter scaling: $\frac{b_0}{\mu} \varepsilon =\mathcal{O}(1)$ as $\varepsilon \to 0^+$.  We introduce the parameter $\eta=\frac{D}{\sigma^2}$ and fix the scaling $\eta =\mathcal{O}(\varepsilon)$ as $ \varepsilon\to 0^+ $.  We further assume that the generating function $ \zeta(\{H\},t) $ is analytic at $ H_i=0 $ for any $ i $ and that the most important contribution to the equation of $ \zeta(\{H\},t) $ comes from a negative real neighborhood of the origin with thickness $ \mathcal{O}(\varepsilon) $. Taking the change of variables $ H_i=\varepsilon S_i $, we expand eq.(\ref{eqzeta_movement}) in powers of $ \varepsilon $, assuming $ S_i=\mathcal{O}(1) $ and $ S_i\leq 0 $. With the definition of the following constants
\[
\lambda=\sqrt{\frac{D}{\mu}} \qquad \rho=\sqrt{\frac{\sigma^2}{b_0}} \quad ,
\]
which are formally equivalent to those defined in section \ref{Mean_and_pair_section}, and retaining only the leading order in $\varepsilon$ at eq.(\ref{eqzeta_movement}), we obtain
\begin{align}
\frac{\partial}{\partial t} \zeta(\{S\},t)=\sum_{i\in \mathbb{L}} \sigma^2 S_i \Big\{ & \eta \Delta_i \frac{\partial \zeta}{\partial S_i}- \varepsilon\frac{\partial \zeta}{\partial S_i}+ \frac{\varepsilon}{\rho^2}\ \zeta +\varepsilon S_i \frac{\partial \zeta}{\partial S_i} \Big\}.
\end{align}
with $\Delta_i$ the discrete Laplace operator. Dividing through by $\varepsilon$ and rescaling time as $T:= \mu t$ we finally obtain
\begin{align}
	\nonumber
	\frac{\partial}{\partial T} \zeta(\{S\},t)=\sum_{i\in \mathbb{L}} S_i \Big\{ & \lambda^2 \Delta_i \frac{\partial \zeta}{\partial S_i}- \frac{\partial \zeta}{\partial S_i}+ \frac{1}{\rho^2}\ \zeta +\\
	&+ S_i \frac{\partial \zeta}{\partial S_i} \Big\} \quad,
\end{align}
which is exactly equal to eq.(\ref{eqzeta_rescaled4}). Since all the results of this paper stem from this equation (or, equivalently, eq.(\ref{eqzeta_rescaled2})), the conclusions drawn in the main text for the model with 'seed dispersal' also hold true for the diffusion model presented in this section, provided the systems are close to the critical point (i.e., as $\varepsilon \to 0$). 

\section{\label{AppE} A null model: the Fisher Log-series}
The Log-series distribution was first proposed in 1943 by the statistician Ronald Fisher to describe the empirical abundance distribution of British moths and Malaysian butterflies \cite{fisher1943relation}. From a set of assumptions involving the independence of species and the absence of spatial dispersal, Fisher derived the following formula, which provides the probability, $P_n$, that a species has $n$ individuals within an ecosystem:
\begin{equation}
\label{eq_fisher}
P_n=-\frac{1}{\log(1-x)} \frac{x^n}{n}
\end{equation}
in which $ n>0 $ and $0<x<1$ is a free parameter that is usually determined via a best-fit to data. The Fisher Log-series is still largely used as a null-model in theoretical ecology \cite{hubbell2001unified,o2010field}, because of the low number of free parameters (only one) and of its straightforward mathematical derivation. Nonetheless, many limitations have emerged \cite{azaele2016statistical,o2018cross} in more recent years. Among these, the lack of a spatial structure has strongly limited its accuracy in describing spatial ecological patterns.

In Fig.\ref{Pasoh_plot} we have compared the predictions of our model to the best-fits of eq.(\ref{eq_fisher}) (rescaled by the total number of species) for three areas of different sizes. The histograms are $\log$-scaled in the $ x $-axis, meaning that the $i$-th bin counts the number of species that have abundances between $e^{i-1}$ and $e^i$. In order to compare these empirical data to the analytical predictions, we first need to compute the probability that a species has abundances between $e^{i-1}$ and $e^i$, which is straightforward from eq.(\ref{spatial_stat_sol}) and from eq.(\ref{eq_fisher}). 

Our model at stationarity has a total of three free parameters (namely, $\langle n \rangle$, $\bar{\lambda}$ and $\bar{\rho}$). $\langle n \rangle$ has been obtained directly from the empirical data, whereas $\bar{\lambda}$ and $\bar{\rho}$ have been calculated from the best fit of the theoretical PCF, i.e. eq.(\ref{pp_correlation}), to the empirical PCF, which is unrelated to the species abundance distribution. Once obtained the three parameters, we have predicted the species abundance distribution by using eq.(\ref{spatial_stat_sol}), which is the curve showed in Fig.(\ref{Pasoh_plot}) along with the histograms of the empirical data.

Looking at the plots in Fig.(\ref{Pasoh_plot}), we see that our model prediction is very accurate in all three cases, whereas the Fisher log-series fails at fitting the distribution in panels (b) and (c). This is confirmed by standard chi-square analysis: $0.91$ $p$-value in the first panel, and $p$-values of $0.7$  in the other two cases; while the Fisher Log-series yields a $p$-value of $0.46$ and $10^{-6}$ in panels (a) and (b), and of $\approx~10^{-21}$ in panel (c). More importantly, our model can predict the SAD at all spatial scales without any further assumptions on the system, whereas a method relying on the fit of eq.(\ref{eq_fisher}) introduces new parameters at each spatial scale (which in general are incompatible with each other) and is inapplicable at scales where we do not have empirical data.

\end{appendices}

\bibliographystyle{unsrt}
\bibliography{bibliography}

\begin{thebibliography}{10}

\bibitem{mora2011biological}
Thierry Mora and William Bialek.
\newblock Are biological systems poised at criticality?
\newblock {\em Journal of Statistical Physics}, 144(2):268--302, 2011.

\bibitem{nykter2008gene}
Matti Nykter, Nathan~D Price, Maximino Aldana, Stephen~A Ramsey, Stuart~A
  Kauffman, Leroy~E Hood, Olli Yli-Harja, and Ilya Shmulevich.
\newblock Gene expression dynamics in the macrophage exhibit criticality.
\newblock {\em Proceedings of the National Academy of Sciences},
  105(6):1897--1900, 2008.

\bibitem{furusawa2012adaptation}
Chikara Furusawa and Kunihiko Kaneko.
\newblock Adaptation to optimal cell growth through self-organized criticality.
\newblock {\em Physical review letters}, 108(20):208103, 2012.

\bibitem{schneidman2006weak}
Elad Schneidman, Michael~J Berry~II, Ronen Segev, and William Bialek.
\newblock Weak pairwise correlations imply strongly correlated network states
  in a neural population.
\newblock {\em Nature}, 440(7087):1007, 2006.

\bibitem{cavagna2010scale}
Andrea Cavagna, Alessio Cimarelli, Irene Giardina, Giorgio Parisi, Raffaele
  Santagati, Fabio Stefanini, and Massimiliano Viale.
\newblock Scale-free correlations in starling flocks.
\newblock {\em Proceedings of the National Academy of Sciences},
  107(26):11865--11870, 2010.

\bibitem{tovo2017upscaling}
Anna Tovo, Samir Suweis, Marco Formentin, Marco Favretti, Igor Volkov,
  Jayanth~R Banavar, Sandro Azaele, and Amos Maritan.
\newblock Upscaling species richness and abundances in tropical forests.
\newblock {\em Science Advances}, 3(10):e1701438, 2017.

\bibitem{athreya2004branching}
Krishna~B Athreya, Peter~E Ney, and PE~Ney.
\newblock {\em Branching processes}.
\newblock Courier Corporation, 2004.

\bibitem{cardy1996scaling}
John Cardy.
\newblock {\em Scaling and renormalization in statistical physics}, volume~5.
\newblock Cambridge university press, 1996.

\bibitem{jahnke2007solving}
Tobias Jahnke and Wilhelm Huisinga.
\newblock Solving the chemical master equation for monomolecular reaction
  systems analytically.
\newblock {\em Journal of mathematical biology}, 54(1):1--26, 2007.

\bibitem{mellor2016characterization}
Andrew Mellor, Mauro Mobilia, and RKP Zia.
\newblock Characterization of the nonequilibrium steady state of a
  heterogeneous nonlinear q-voter model with zealotry.
\newblock {\em EPL (Europhysics Letters)}, 113(4):48001, 2016.

\bibitem{pavliotis2014stochastic}
Grigorios~A Pavliotis.
\newblock {\em Stochastic processes and applications: diffusion processes, the
  Fokker-Planck and Langevin equations}, volume~60.
\newblock Springer, 2014.

\bibitem{zia2007probability}
RKP Zia and B~Schmittmann.
\newblock Probability currents as principal characteristics in the statistical
  mechanics of non-equilibrium steady states.
\newblock {\em Journal of Statistical Mechanics: Theory and Experiment},
  2007(07):P07012, 2007.

\bibitem{henkel2008non}
Malte Henkel, Haye Hinrichsen, and Sven L{\"u}beck.
\newblock {\em Non-Equilibrium Phase Transitions: Volume 1: Absorbing Phase
  Transitions}.
\newblock Springer Science \& Business Media, 2008.

\bibitem{garcia2012noise}
Jordi Garc{\'\i}a-Ojalvo and Jos{\'e} Sancho.
\newblock {\em Noise in spatially extended systems}.
\newblock Springer Science \& Business Media, 2012.

\bibitem{shem2017solution}
Yahav Shem-Tov, Matan Danino, and Nadav~M Shnerb.
\newblock Solution of the spatial neutral model yields new bounds on the
  amazonian species richness.
\newblock {\em Scientific reports}, 7:42415, 2017.

\bibitem{krapivsky2010kinetic}
Pavel~L Krapivsky, Sidney Redner, and Eli Ben-Naim.
\newblock {\em A kinetic view of statistical physics}.
\newblock Cambridge University Press, 2010.

\bibitem{grilli2012absence}
Jacopo Grilli, Sandro Azaele, Jayanth~R Banavar, and Amos Maritan.
\newblock Absence of detailed balance in ecology.
\newblock {\em EPL (Europhysics Letters)}, 100(3):38002, 2012.

\bibitem{o2010field}
James~P O'Dwyer and Jessica~L Green.
\newblock Field theory for biogeography: a spatially explicit model for
  predicting patterns of biodiversity.
\newblock {\em Ecology letters}, 13(1):87--95, 2010.

\bibitem{o2018cross}
James~P O’Dwyer and Stephen~J Cornell.
\newblock Cross-scale neutral ecology and the maintenance of biodiversity.
\newblock {\em Scientific reports}, 8(1):10200, 2018.

\bibitem{azaele2016statistical}
Sandro Azaele, Samir Suweis, Jacopo Grilli, Igor Volkov, Jayanth~R Banavar, and
  Amos Maritan.
\newblock Statistical mechanics of ecological systems: Neutral theory and
  beyond.
\newblock {\em Reviews of Modern Physics}, 88(3):035003, 2016.

\bibitem{lebedev2012special}
N.N. Lebedev and R.A. Silverman.
\newblock {\em Special Functions \& Their Applications}.
\newblock Dover Books on Mathematics. Dover Publications, 2012.

\bibitem{gardiner2004handbook}
C.W. Gardiner.
\newblock {\em Handbook of Stochastic Methods for Physics, Chemistry, and the
  Natural Sciences}.
\newblock Springer complexity. Springer, 2004.

\bibitem{van1992stochastic}
N.~G. Van~Kampen.
\newblock {\em Stochastic processes in physics and chemistry}, volume~1.
\newblock Elsevier, 1992.

\bibitem{Volkov2005}
I~Volkov, JR~Banavar, F~He, S~Hubbell, and A~Maritan.
\newblock {Density dependence explains tree species abundance and diversity in
  tropical forests.}
\newblock {\em Nature}, 438(7068):658--61, December 2005.

\bibitem{Volkov2007}
I~Volkov, J~R Banavar, S~P Hubbell, and A~Maritan.
\newblock {Patterns of relative species abundance in rainforests and coral
  reefs}.
\newblock {\em Nature}, 450(7166):45--49, 2007.

\bibitem{taylor1961aggregation}
LR~Taylor.
\newblock Aggregation, variance and the mean.
\newblock {\em Nature}, 189(4766):732, 1961.

\bibitem{taylor1977aggregation}
LR~Taylor and RAJ Taylor.
\newblock Aggregation, migration and population mechanics.
\newblock {\em Nature}, 265(5593):415, 1977.

\bibitem{eisler2008fluctuation}
Zolt{\'a}n Eisler, Imre Bartos, and J{\'a}nos Kert{\'e}sz.
\newblock Fluctuation scaling in complex systems: Taylor's law and beyond.
\newblock {\em Advances in Physics}, 57(1):89--142, 2008.

\bibitem{james2018zipf}
Charlotte James, Sandro Azaele, Amos Maritan, and Filippo Simini.
\newblock Zipf's and taylor's laws.
\newblock {\em Physical Review E}, 98(3):032408, 2018.

\bibitem{r01748}
T.~Kolokotrones, V.~Savage, E.~J. Deeds, and W.~Fontana.
\newblock Curvature in metabolic scaling.
\newblock {\em Nature}, 464:753{\textendash}756, 2010.
\newblock [News \&amp; Views feature by Craig R. White, "There is no single p",
  Nature, 464:691 (2010)].

\bibitem{Zaoli17323}
Silvia Zaoli, Andrea Giometto, Emilio Mara{\~n}{\'o}n, St{\'e}phane Escrig,
  Anders Meibom, Arti Ahluwalia, Roman Stocker, Amos Maritan, and Andrea
  Rinaldo.
\newblock Generalized size scaling of metabolic rates based on single-cell
  measurements with freshwater phytoplankton.
\newblock {\em Proceedings of the National Academy of Sciences},
  116(35):17323--17329, 2019.

\bibitem{azaele2006dynamical}
Sandro Azaele, Simone Pigolotti, Jayanth~R Banavar, and Amos Maritan.
\newblock Dynamical evolution of ecosystems.
\newblock {\em Nature}, 444(7121):926, 2006.

\bibitem{Gillespie1977exact}
Daniel~T Gillespie.
\newblock Exact stochastic simulation of coupled chemical reactions.
\newblock {\em The journal of physical chemistry}, 81(25):2340--2361, 1977.

\bibitem{peruzzo2017phenomenological}
Fabio Peruzzo and Sandro Azaele.
\newblock A phenomenological spatial model for macro-ecological patterns in
  species-rich ecosystems.
\newblock In {\em Stochastic Processes, Multiscale Modeling, and Numerical
  Methods for Computational Cellular Biology}, pages 349--368. Springer, 2017.

\bibitem{Dornic2005integration}
Ivan Dornic, Hugues Chat{\'e}, and Miguel~A Munoz.
\newblock Integration of langevin equations with multiplicative noise and the
  viability of field theories for absorbing phase transitions.
\newblock {\em Physical review letters}, 94(10):100601, 2005.

\bibitem{condit2002beta}
Richard Condit, Nigel Pitman, Egbert~G Leigh, J{\'e}r{\^o}me Chave, John
  Terborgh, Robin~B Foster, Percy N{\'u}nez, Salom{\'o}n Aguilar, Renato
  Valencia, Gorky Villa, et~al.
\newblock Beta-diversity in tropical forest trees.
\newblock {\em Science}, 295(5555):666--669, 2002.

\bibitem{chave2002spatially}
Jer{\^o}me Chave and Egbert~G Leigh~Jr.
\newblock A spatially explicit neutral model of $\beta$-diversity in tropical
  forests.
\newblock {\em Theoretical population biology}, 62(2):153--168, 2002.

\bibitem{coleman1981random}
Bernard~D Coleman.
\newblock On random placement and species-area relations.
\newblock {\em Mathematical Biosciences}, 54(3-4):191--215, 1981.

\bibitem{mcgill2007species}
Brian~J McGill, Rampal~S Etienne, John~S Gray, David Alonso, Marti~J Anderson,
  Habtamu~Kassa Benecha, Maria Dornelas, Brian~J Enquist, Jessica~L Green,
  Fangliang He, et~al.
\newblock Species abundance distributions: moving beyond single prediction
  theories to integration within an ecological framework.
\newblock {\em Ecology letters}, 10(10):995--1015, 2007.

\bibitem{black2012stochastic}
Andrew~J Black and Alan~J McKane.
\newblock Stochastic formulation of ecological models and their applications.
\newblock {\em Trends in ecology \& evolution}, 27(6):337--345, 2012.

\bibitem{pigolotti2018stochastic}
Simone Pigolotti, Massimo Cencini, Daniel Molina, and Miguel~A Mu{\~n}oz.
\newblock Stochastic spatial models in ecology: a statistical physics approach.
\newblock {\em Journal of Statistical Physics}, 172(1):44--73, 2018.

\bibitem{hubbell2001unified}
Stephen~P Hubbell.
\newblock {\em The unified neutral theory of biodiversity and biogeography
  (MPB-32)}.
\newblock Princeton University Press, 2001.

\bibitem{rinaldo2002cross}
Andrea Rinaldo, Amos Maritan, Kent~K Cavender-Bares, and Sallie~W Chisholm.
\newblock Cross--scale ecological dynamics and microbial size spectra in marine
  ecosystems.
\newblock {\em Proceedings of the Royal Society of London. Series B: Biological
  Sciences}, 269(1504):2051--2059, 2002.

\bibitem{ser2018ubiquitous}
Enrico Ser-Giacomi, Lucie Zinger, Shruti Malviya, Colomban De~Vargas, Eric
  Karsenti, Chris Bowler, and Silvia De~Monte.
\newblock Ubiquitous abundance distribution of non-dominant plankton across the
  global ocean.
\newblock {\em Nature ecology \& evolution}, 2(8):1243, 2018.

\bibitem{woodcock2007neutral}
Stephen Woodcock, Christopher~J Van Der~Gast, Thomas Bell, Mary Lunn, Thomas~P
  Curtis, Ian~M Head, and William~T Sloan.
\newblock Neutral assembly of bacterial communities.
\newblock {\em FEMS microbiology ecology}, 62(2):171--180, 2007.

\bibitem{giometto2014emerging}
Andrea Giometto, Andrea Rinaldo, Francesco Carrara, and Florian Altermatt.
\newblock Emerging predictable features of replicated biological invasion
  fronts.
\newblock {\em Proceedings of the National Academy of Sciences},
  111(1):297--301, 2014.

\bibitem{giometto2015sample}
Andrea Giometto, Marco Formentin, Andrea Rinaldo, Joel~E Cohen, and Amos
  Maritan.
\newblock Sample and population exponents of generalized taylor’s law.
\newblock {\em Proceedings of the National Academy of Sciences},
  112(25):7755--7760, 2015.

\bibitem{hanson2012beyond}
China~A Hanson, Jed~A Fuhrman, M~Claire Horner-Devine, and Jennifer~BH Martiny.
\newblock Beyond biogeographic patterns: processes shaping the microbial
  landscape.
\newblock {\em Nature Reviews Microbiology}, 10(7):497, 2012.

\bibitem{altermatt2015big}
Florian Altermatt, Emanuel~A Fronhofer, Aurelie Garnier, Andrea Giometto,
  Frederik Hammes, Jan Klecka, Delphine Legrand, Elvira M{\"a}chler, Thomas~M
  Massie, Frank Pennekamp, et~al.
\newblock Big answers from small worlds: a user's guide for protist microcosms
  as a model system in ecology and evolution.
\newblock {\em Methods in Ecology and Evolution}, 6(2):218--231, 2015.

\bibitem{lynch2015ecology}
Michael~DJ Lynch and Josh~D Neufeld.
\newblock Ecology and exploration of the rare biosphere.
\newblock {\em Nature Reviews Microbiology}, 13(4):217, 2015.

\bibitem{azaele2015towards}
Sandro Azaele, Amos Maritan, Stephen~J Cornell, Samir Suweis, Jayanth~R
  Banavar, Doreen Gabriel, and William~E Kunin.
\newblock Towards a unified descriptive theory for spatial ecology: predicting
  biodiversity patterns across spatial scales.
\newblock {\em Methods in Ecology and Evolution}, 6(3):324--332, 2015.

\bibitem{mastromatteo2011criticality}
Iacopo Mastromatteo and Matteo Marsili.
\newblock On the criticality of inferred models.
\newblock {\em Journal of Statistical Mechanics: Theory and Experiment},
  2011(10):P10012, 2011.

\bibitem{goudarzi2012emergent}
Alireza Goudarzi, Christof Teuscher, Natali Gulbahce, and Thimo Rohlf.
\newblock Emergent criticality through adaptive information processing in
  boolean networks.
\newblock {\em Physical review letters}, 108(12):128702, 2012.

\bibitem{stieg2012emergent}
Adam~Z Stieg, Audrius~V Avizienis, Henry~O Sillin, Cristina Martin-Olmos,
  Masakazu Aono, and James~K Gimzewski.
\newblock Emergent criticality in complex turing b-type atomic switch networks.
\newblock {\em Advanced Materials}, 24(2):286--293, 2012.

\bibitem{hidalgo2014information}
Jorge Hidalgo, Jacopo Grilli, Samir Suweis, Miguel~A Mu{\~n}oz, Jayanth~R
  Banavar, and Amos Maritan.
\newblock Information-based fitness and the emergence of criticality in living
  systems.
\newblock {\em Proceedings of the National Academy of Sciences},
  111(28):10095--10100, 2014.

\bibitem{gnesotto2018broken}
FS~Gnesotto, Federica Mura, Jannes Gladrow, and CP~Broedersz.
\newblock Broken detailed balance and non-equilibrium dynamics in living
  systems: a review.
\newblock {\em Reports on Progress in Physics}, 81(6):066601, 2018.

\bibitem{battle2016broken}
Christopher Battle, Chase~P Broedersz, Nikta Fakhri, Veikko~F Geyer, Jonathon
  Howard, Christoph~F Schmidt, and Fred~C MacKintosh.
\newblock Broken detailed balance at mesoscopic scales in active biological
  systems.
\newblock {\em Science}, 352(6285):604--607, 2016.

\bibitem{abramowitz1965handbook}
Milton Abramowitz and Irene~A Stegun.
\newblock {\em Handbook of mathematical functions: with formulas, graphs, and
  mathematical tables}, volume~55.
\newblock Courier Corporation, 1965.

\bibitem{fisher1943relation}
Ronald~A Fisher, A~Steven Corbet, and Carrington~B Williams.
\newblock The relation between the number of species and the number of
  individuals in a random sample of an animal population.
\newblock {\em The Journal of Animal Ecology}, pages 42--58, 1943.

\end{thebibliography}

\end{document}